\documentclass[aps,twocolumn,showpacs,prl,amsmath,amssymb,floatfix,superscriptaddress]{revtex4-2}
\usepackage[utf8]{inputenc}
\usepackage[usenames,dvipsnames]{xcolor}
\usepackage{amsmath}
\usepackage{mathtools}
\usepackage{graphicx}
\usepackage{esint}
\usepackage{appendix}
\usepackage{hyperref}
\usepackage{comment}

\usepackage[normalem]{ulem} 

\providecommand{\openone}{\mathbf{1}}

\definecolor{pdarkblue}{rgb}{0.1797, 0.1875, 0.5703}
\usepackage{hyperref}
\hypersetup{
    colorlinks = true,
    citecolor = pdarkblue,
    linkcolor = pdarkblue,
    urlcolor = pdarkblue,
}

\makeatletter
\@ifundefined{textcolor}{}
{%
 \definecolor{BLACK}{gray}{0}
 \definecolor{WHITE}{gray}{1}
 \definecolor{RED}{rgb}{1,0,0}
 \definecolor{GREEN}{rgb}{0,1,0}
 \definecolor{BLUE}{rgb}{0,0,1}
 \definecolor{CYAN}{cmyk}{1,0,0,0}
 \definecolor{MAGENTA}{cmyk}{0,1,0,0}
 \definecolor{YELLOW}{cmyk}{0,0,1,0}
}
\makeatother

\allowdisplaybreaks

\begin{document}
\title{Random Coarse-Graining with Applications to Measuring Irreversibility 
}
\author{Ruicheng Bao}
\thanks{Corresponding author: ruicheng@g.ecc.u-tokyo.ac.jp}
\affiliation{Department of Physics, Graduate School of Science, The University of Tokyo, 7-3-1 Hongo, Bunkyo-ku, Tokyo 113-0033, Japan}
\author{Naruo Ohga}
\affiliation{Department of Physics, Graduate School of Science, The University of Tokyo, 7-3-1 Hongo, Bunkyo-ku, Tokyo 113-0033, Japan}
\author{Sosuke Ito}
\affiliation{Department of Physics, Graduate School of Science, The University of Tokyo, 7-3-1 Hongo, Bunkyo-ku, Tokyo 113-0033, Japan}
\affiliation{Universal Biology Institute, Graduate School of Science, The University of Tokyo, 7-3-1 Hongo, Bunkyo-ku, Tokyo 113-0033, Japan}

\date{\today}
\begin{abstract}
Thermodynamic irreversibility is a fundamental concept in statistical physics, yet its experimental measurement remains challenging, especially for complex systems. {At the same time, imprecise
readout is ubiquitous in experiments, but its effect on thermodynamic irreversibility has not
been formulated in a general way.} We introduce a novel random coarse-graining framework that incorporates a probabilistic mapping from fine-grained to coarse-grained states, and we use it {both to model imprecise measurements and} to identify model-free measures of irreversibility in complex many-body systems. These measures are constructed from the asymmetry of cross-correlation functions between suitably chosen observables, providing rigorous lower bounds on the entropy production. For many-particle systems, we propose a particularly practical implementation that divides real space into virtual boxes and monitors particle number densities within them, requiring only simple counting from video microscopy, without single-particle tracking, trajectory reconstruction, or prior knowledge of interactions. Owing to its generality and {limited} data requirements, the random coarse-graining framework offers broad applicability across diverse nonequilibrium systems.

\end{abstract}
\maketitle

\textit{Introduction}---Thermodynamic irreversibility, which manifests itself as the ``arrow of time'', stands as a cornerstone concept in statistical physics. Its significance extends across diverse fields, ranging from heat engines \cite{esposito09prl,esposito10prl,shiraishi16tradeoff,pietzonka18prl} and other thermal machines \cite{seifert11prl,pietzonka2016universal,seifert19prx} to biological systems \cite{wang19RMP,ito2015maxwell,seifert17coherence,seifert22coherent,23ohgaprl,24artemyprr,24itoPRX}, chemical reaction networks \cite{esposito16prx,circuit2022prx,polettini22prx,crn24prl,rds24prl,liang2024thermodynamic}, electrical networks \cite{electric20prx} and even computer algorithm optimization \cite{24PRXWolpert,wolpert2024stochastic}.  Measuring irreversibility can provide valuable insights for these domains, for instance, enabling the optimization of heat engines \cite{esposito09prl,esposito10prl,shiraishi16tradeoff,pietzonka18prl}, nano-machines \cite{seifert11prl,pietzonka2016universal} and energy-transduction processes \cite{di2024variance}, while deepening our understanding of how biological systems maintain their complex structures and functions \cite{24PRXLifeirre}. 

Entropy production (EP) is a widely utilized measure of irreversibility, largely due to its intrinsic connection with heat dissipation. Despite its importance, measuring EP poses significant challenges, even in relatively simple systems. Recent years have seen a surge in thermodynamic inference research aimed at addressing this issue \cite{Seifert19inference,martinez_inferring_2019,20itoprx,skinner_estimating_2021,22PRL_EPmeasure,MES22PRX,HDPR22PRX,VDS23prl,blom_milestoning_2024,pedro24pre,2024seifertpnas,24seifertpre_infer,24seifertprr_infer,24raminprr,knotz2024fphy,24ohgaprr,di2024variance,meyberg_entropy_2024,bao2024nonequilibrium,Jann25faulty,Auconi_25prl}. {However, most existing inference schemes
implicitly assume that the accessible states or observables are specified without measurement
error. Additionally, many of these approaches have been developed and benchmarked mainly in relatively simple Markov-state or low-dimensional Langevin settings.} While the thermodynamic uncertainty relation (TUR) \cite{15PRL_TUR,16PRL_TUR,17PRE_finiteTUR,17PRE_SeifertTUR,dechant2018current,dechant2018multidimensional,dechant2020fluctuation,19PRE_CRITUR,21PRX_ImprovedTUR,20PRL_Short,20PRE_shortTUR,20uedaTUR,seiferttimeTUR,seifert_interactingTUR,23godectur,23Baopre} offers viable lower bounds through current statistics, measuring such statistics is typically hard in many-body systems \cite{Arnab21TUR}. Indeed, TURs are primarily applied to simple systems with few exceptions, as in \cite{Arnab21TUR,seifert_interactingTUR,miguel25inferring} for specific cases. Moreover, the primary challenge of experimental measurements lies in reconstructing trajectory information, a process that is often complex and resource-intensive, especially in dense and heterogeneous systems where particle tracking becomes difficult \cite{count24prx}. In such cases, even the TUR may be inapplicable, as it requires precise positional information of tagged particles and relies on homogeneity \cite{seifert_interactingTUR}. {Therefore, a comprehensive framework that can
both measure irreversibility in complex many-body systems and systematically incorporate imprecise
measurements remains elusive but is essential for both practical applications and fundamental
understanding.} 
 
In this Letter, we bridge this gap by developing a systematic coarse-graining framework, which redefines the conventional wisdom in nonequilibrium physics that coarse-graining procedures should be many-to-one mappings \cite{broeck07dissipation,esposito12pre,Seifert19inference,MES22PRX,HDPR22PRX,VDS23prl}. Our new framework, termed random coarse-graining, incorporates many-to-many mappings, allowing for overlaps between coarse-grained (CG) states or trajectories. This approach offers clear advantages over existing coarse-graining schemes, which are often difficult to apply universally to interacting many-body systems \cite{voth2022bottom}. Traditional mappings also present difficulties in flexibly adjusting the degree of coarse-graining, with high degrees of reduction hindering the inference of thermodynamic quantities due to information loss. 

Within this framework, { we clarify how measurement errors affect the inferable entropy production and} propose experimentally accessible measures of irreversibility that can be obtained from the asymmetry of cross-correlation functions between suitably chosen observables, applicable to both nonequilibrium steady states and arbitrary non-stationary processes, and requiring neither trajectory reconstruction nor detailed knowledge of the dynamics. These measures serve as lower bounds on the EP, acting as reliable irreversibility indicators that properly vanish for equilibrium systems. {In the many-body setting, a particularly practical choice of observables is the particle number density in artificially divided spatial regions (virtual boxes), whose measurement demands only {very coarse counting information} and does not require single-particle tracking.} Since particle number is a generic and routinely measured observable in many-body systems, our approach is broadly applicable and facilitates experimental realization. Additionally, as these measures consist of positive contributions from distinct spatial regions, they enable a natural decomposition of irreversibility into local contributions. Unlike conventional coarse-graining methods, this decomposition characterizes spatial distribution of irreversibility beyond a single global metric. The framework also offers flexibility in adjusting the degree of coarse-graining by modifying the number of virtual boxes. 

\textit{Setup.}--- We consider a system whose underlying fine-grained (FG) dynamics (i.e., microscopic dynamics) can be described by a master equation:
\begin{equation}
    \frac{d}{dt} p_{i}(t) = \sum_j \left[k_{ij}(t) p_j(t)-k_{ji}(t) p_i(t)\right], \label{ME}
\end{equation}
where \( p_i(t) \) denotes the probability of finding the system in state \( i \) at time \( t \), and \( k_{ij}(t) \) is the transition rate from state \( j \) to state \( i \) at time \( t \), satisfying $\sum_j k_{ji}(t)=0$ by conservation of probability. The time dependence of the transition rates can encode arbitrary driving protocols. The system is connected with a single or multiple heat baths. With multiple baths, transition rates are composed of contributions from different baths as $k_{ij}(t)=\sum_{\nu}k_{ij}^{\nu}(t)$, where $k_{ij}^{\nu}(t)$ is the transition rate associated with the $\nu$-th bath. {This channel-resolved decomposition is exact when each elementary transition can be assigned to a bath label $\nu$. If the bath or process label is itself unobserved and is coarse-grained over, additional hidden dissipation can appear \cite{Busiello20prr}.} Assuming that each bath obeys the local detailed balance, the EP rate (EPR) at time \( t \) is given by \cite{76RMP, seifert2012stochastic}:
\begin{equation}
\dot{\sigma}(t) \coloneqq  
\sum_{\nu}\sum_{i,j} k^{\nu}_{ij}(t)p_j(t)\ln\frac{k^{\nu}_{ij}(t)p_j(t)}{k^{\nu}_{ji}(t)p_i(t)},
\end{equation}
where we set the Boltzmann constant to unity.
Then, the EP over the interval {\([t, t+\tau]\) is \(\sigma_{[t, t+\tau]} \coloneqq  \int_t^{t+\tau} \dot{\sigma}(t)\, dt\).} 

By appropriately taking the continuum limit of the state space (phase space), Eq.~\eqref{ME} can encompass many-body Langevin dynamics with arbitrary confining potentials and interactions \cite{shiraishi16tradeoff,shiraishi2023introduction}. The EP remains invariant under this continuum limit \cite{shiraishi16tradeoff,76RMP}, ensuring that our subsequent results extend smoothly to systems governed by many-body Langevin dynamics.

\textit{Effective master equation for random coarse-graining dynamics.}--- In previous studies \textcolor{black}{that used deterministic lumping} (e.g., see ~\cite{esposito12pre,hilder2024quantitative}), coarse-graining is usually defined by summing the probabilities of FG states to obtain the probability of a single CG state $m$: $\sum_{i\in S_{m}} p_i $, where $S_m$ is the set of FG states comprising the CG state $m$. { Note that $\{S_m\}$ ($m=1,2,...$) should be a partition of the total state space.}   

\begin{figure}
    \centering
    \includegraphics[width=1.0\linewidth]{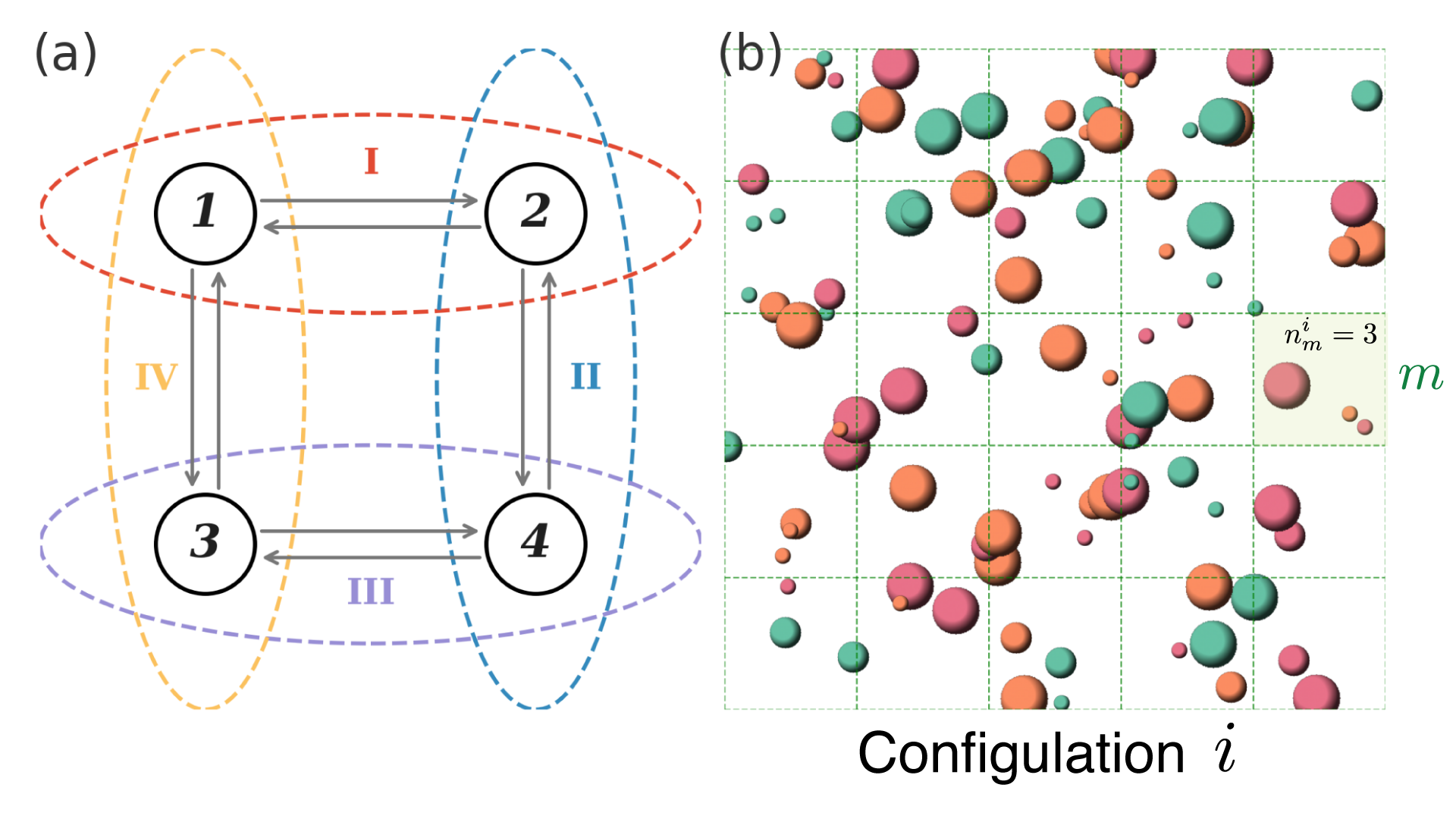}
    \caption{Illustration of our coarse-graining method. (a) A four-state unicycle model. \textcolor{black}{The FG states are $1$, $2$, $3$ and $4$. The CG version of the model still has four CG states, $\rm I$, $\rm II$, $\rm III$ and $\rm IV$.} \textcolor{black}{For example,} the conditional probability that state $1$ is in state I (IV) is $T (\rm I|1)$ ($T (\rm IV|1)$), and $T (\rm I|1)+T (\rm IV|1)=1$ is satisfied. (b) {A snapshot of microscopic configuration $i$ of a many-particle system} with heterogeneous particle radii and charges (represented by size and color, respectively). The real space is divided by virtual boxes (green dotted line). Here, we use $T (m|i)= \rho_m^i := {n_m^i}/{n_{\rm tot}}$ as the conditional probability, where {$n_m^i$ is the particle number in box $m$ for configuration $i$}. \textcolor{black}{The CG distribution $p^{\rm cg}_m$ reads $\langle\rho_m (t) \rangle := \sum_i \rho_m^i p_i(t)$, {where $\langle\cdot\rangle$ denotes the ensemble average over FG states.}}}
    \label{1}
\end{figure}
Here, we generalize this by introducing a random mapping from a FG state $i$ to a CG state $m$ by a conditional probability $T (m|i)\geq 0$ with $\sum_m T (m|i)=1$. {This kernel is not explicitly time-dependent: all time dependence enters through the evolving FG state and distribution. By the law of total probability,} the probability of the system being in a CG state $m$ is 
\begin{equation}
    p^{\rm cg}_{m}(t) :=\sum_i T (m|i) p_i(t).\label{cgmapping}
\end{equation}
An illustrative example of this coarse graining is shown in Fig.~\ref{1}(a). From the master equation \eqref{ME} and the definition \eqref{cgmapping}, we first derive an effective master equation for the CG dynamics as (see the End Matter)
\begin{equation}
    \frac{d}{dt} p^{\rm cg}_{m}(t) = \sum_{n} \left[k^{\rm cg}_{mn}(t)  p^{\rm cg}_n(t)-k^{\rm cg}_{nm}(t)  p^{\rm cg}_m(t)\right], \label{EME}
\end{equation}
where the effective transition rate ($n\neq m$) $k^{\rm cg}_{mn}(t) \coloneqq \sum_{i,j|i\neq j} T (m|i)k_{ij}(t) T (n|j) p_j(t) /p^{\rm cg}_n (t) \geq 0 $ is time-dependent even when the FG transition rates are time-independent. Diagonal elements $k_{mm}^{\rm cg}(t)$ are naturally defined so that the conservation of probability $\sum_m k_{mn}^{\rm cg}(t) = 0$ is satisfied. 

Equation~\eqref{EME} reduces to the effective master equation for deterministic lumping when choosing the mapping $T (m|i)$ as indicator functions of the CG states, i.e., $T (m|i)=\openone_{S_m} (i)$. Here, $\openone_{A} (i)=1\ \text{if}\ i\in A, \text{and}\ \openone_{A} (i) =0\ \text{if}\ i\notin A$. \textcolor{black}{Because $S_m$ is a partition of the total state space, $\sum_m T(m|i)=1$ is satisfied. Therefore, our framework generalizes the results of deterministic lumping.}

With this effective equation, we prove that the EPR defined at the CG level is a lower bound of the true EPR:
\begin{equation}
    \dot{\sigma}(t) \geq \dot{\sigma}^{\rm cg}(t)\coloneqq  \sum_{m,n} k^{\rm cg}_{mn} p^{\rm cg}_n \ln{\frac{k^{\rm cg}_{mn} p^{\rm cg}_n}{k^{\rm cg}_{nm} p^{\rm cg}_m}}\geq 0.
    \label{lowerbound}
\end{equation}
Further, the EPR can be decomposed into three non-negative components: 
\begin{equation}
    \dot{\sigma}(t) = \dot{\sigma}^{\rm cg}(t) + \dot{\sigma}^{\text{inn}}(t)+\dot{\sigma}^{\text{tran}}(t),
\end{equation}
where 
\begin{equation}
\dot{\sigma}^{\text{inn}} \coloneqq \sum_m \sum_\nu \sum_{i,j|i\neq j}T (m|i)T (m|j) k^\nu_{ij} p_j \ln \frac{k^\nu_{ij} p_j}{k^\nu_{ji} p_i} \geq 0
\end{equation}
is the hidden EPR within each CG state, and {$\dot{\sigma}^{\text{tran}}(t):= \dot{\sigma}(t) -\dot{\sigma}^{\rm cg}(t) - \dot{\sigma}^{\text{inn}}(t)$ can be regarded as the hidden EPR due to coarse-graining of transition rates. We prove $\dot{\sigma}^{\text{tran}}(t)\geq 0$ in the End Matter.} The decomposition generalizes the previous result on stochastic thermodynamics under lumping \cite{esposito12pre} to random lumping and clarifies the physical meanings of the contributions dropped in the inequality \eqref{lowerbound}.

\begin{figure}
    \centering
    \includegraphics[width=0.7\linewidth]{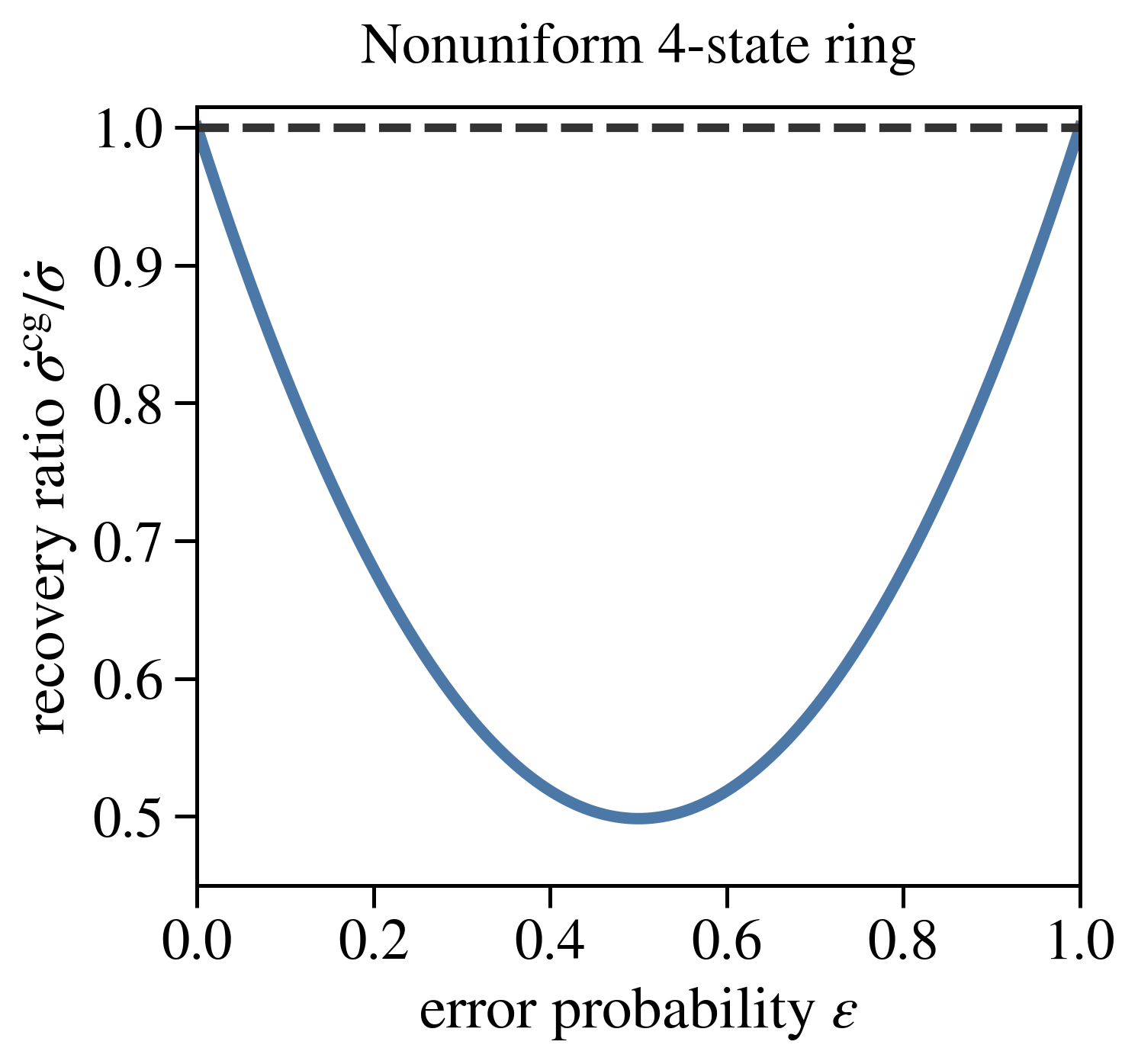}
    \caption{{ Recovery ratio $\dot{\sigma}^{\rm cg}/\dot{\sigma}$ versus the error probability $\varepsilon$ for a nonuniform four-state ring. 
    }}
    \label{fig:imprecise}
\end{figure}

{Our general framework in Eqs.~\eqref{cgmapping}--\eqref{lowerbound} admits multiple physical interpretations.

\textit{Impact of imprecise measurement on thermodynamics---}First, it provides a natural framework for the stochastic thermodynamics of imprecise measurements. In many experiments, the microscopic state $i$ cannot be identified perfectly; instead, the detector returns one among several outcomes $m$, with an error probability that depends on the underlying FG state. Within our framework, this measurement imperfection is described precisely by the random mapping $T(m|i)$: for a fixed FG state $i$, the apparatus reports the outcome $m$ only probabilistically. The experimentally accessible distribution is therefore the distribution of measurement outcomes, $p_m^{\rm cg}(t)=\sum_i T(m|i)p_i(t)$, rather than the FG distribution itself. Different FG states can be confused with each other, and even the same FG state can lead to different readouts in repeated measurements, reflecting finite resolution or measurement noise. The FG dynamics is recovered as the error-free limit in which $T(m|i)$ becomes an indicator function and each FG state is assigned to a unique outcome. For simplicity, we illustrate this idea using a one-parameter nearest-neighbor measurement-error channel on a uniform four-state ring ($i \in \{1,2,3,4\}$ and $m \in \{1,2,3,4\}$),
\begin{equation}
    T(m|i)=(1-\varepsilon)\delta_{m,i}+\varepsilon\,\delta_{m,i-1},
\end{equation}
where $\varepsilon\in[0,1]$ is the readout-error probability and interpret the index $i-1$ modulo $4$ (see Fig. \ref{1}(a) for illustration and \cite{supp_mat} Sec. II for details).  
In this example, $\dot{\sigma}^{\rm cg}/\dot{\sigma}=(1-\varepsilon)^2+\varepsilon^2$, so the inferable EP decreases monotonically with $\varepsilon$ and reaches its minimum $\dot{\sigma}^{\rm cg}=\dot{\sigma}/2$ at $\varepsilon=1/2$. A nonuniform ring also shows similar behaviors; see Figure~\ref{fig:imprecise}. We prove in \cite{supp_mat} that this monotonic decrease (increase) on $0\leq \varepsilon\leq 1/2$ ($1/2\le \varepsilon \le 1$) persists for a broader class of single-error-probability channels with pairwise adjacent confusion. Interestingly, there exist exact-recovery families with $\dot{\sigma}^{\rm cg}=\dot{\sigma}$ for all $\varepsilon$. We provide such an example in Sec. II of \cite{supp_mat}.} 

{\textit{Estimating irreversibility via cross-correlations between arbitrary observables---}Another, more operational, interpretation is to regard $T(m|i)$ as a set of experimentally accessible observables. Consider measuring a collection of non-negative observables $O_1'(t), O_2'(t),\dots,O_M'(t)$. If the system happens to be in the FG state $i$ at time $t$, then a single measurement returns the output vector $\{O_1^{\prime\,i}, O_2^{\prime\,i}, \dots\}$. We then normalize this single-shot output by its total intensity, defining $O_m^i\equiv {O'}_m^{i}/{O'}^{i}_{\rm tot}$ with ${O'}^{i}_{\rm tot} \coloneqq \sum_m {O'}_m^{i}$, so that $\sum_m O_m^i=1$. We here assume that ${O'}^{i}_{\rm tot}>0$. This normalization uses only the measured output vector and does not require any additional measurement or any knowledge of the hidden state. Identifying $T(m|i)=O_m^i$ therefore means that each FG state is associated with a probability vector over effective outcomes. 
}
Under this interpretation, we can introduce the cross-correlation functions
\begin{align}
    C_{nm}^{t,\tau} &\equiv \langle O_n(t) O_m (t+\tau) \rangle \coloneqq 
    \sum_{i,j} \,O_n^iO_m^j \mathcal{P}[j,t+\tau;i,t],
    \notag \\
    B_{n m}^{t,\tau} &\coloneqq \sum_{i} O_n^iO_m^i \mathcal{P}[i,t+\tau;i,t],
    \label{corr-def}
\end{align}
where $\mathcal{P}[j,t+\tau;i,t]$ is the joint probability of being in $i$ at time $t$ and in $j$ at time $t+\tau$. The function $C_{nm}^{t,\tau}$ is the physical (measurable) correlation function, and $B_{nm}^{t,\tau}$ is the correlation averaged over realizations that keep the FG state invariant. These definitions allow us to express the CG EPR as
\begin{equation}
    \label{CG-EPR-corr}
    \dot{\sigma} \geq \dot{\sigma}^{\rm cg}= \lim_{\Delta t\to 0} \!\sum_{m,n|m> n} \!\!\frac{C^{t,\Delta t}_{nm} - C^{t,\Delta t}_{mn}}{\Delta t} \ln \frac{C^{t,\Delta t}_{nm} - B^{t,\Delta t}_{nm}}{C^{t,\Delta t}_{mn} - B^{t,\Delta t}_{mn}},
\end{equation}
as shown in the End Matter. Omitting $B^{t,\Delta t}_{mn}$ also yields a weaker bound 
\begin{align}
    \dot{\sigma} \geq \dot{\sigma}^{\rm cg} \geq \lim_{\Delta t\to 0} \sum_{m,n} \frac{C^{t,\Delta t}_{nm} }{\Delta t} \ln \frac{C^{t,\Delta t}_{nm}}{C^{t,\Delta t}_{mn}}.
    \label{bound-c}
\end{align}
These lower bounds of the true EP serve as measures of irreversibility that are experimentally accessible via the correlation functions.

The infinitesimal-time framework developed above can be extended to a finite time evolution from $t$ to $t +\tau$:
\begin{equation}
   \sigma_{[t,t+\tau]} 
    \geq \sum_{m,n} C^{t,\tau}_{nm} \ln{\frac{C^{t,\tau}_{nm}}{C^{\dagger, t,\tau}_{mn}}}
   \eqqcolon \sigma_{[t,t+\tau]}^{\rm est},\label{finite2}
\end{equation}
where $C^{\dagger, t,\tau}_{mn}$ is the cross-correlation function defined via the backward probability $\mathcal{P}^{\dagger}[i,t+\tau;j,t]=\mathcal{P}^{\dagger}[i,t+\tau|j,t]p_j(t+\tau)$ {(see the End Matter for details)}. This lower bound can be viewed as a finite-time generalization of Eq.~\eqref{bound-c}. { The inequality follows from an application of the log-sum inequality 
to the finite-time propagators.}

Beyond its practical utility for inference, $\sigma_{[t,t+\tau]}^{\rm est}$ has a clear physical interpretation: it is a normalized measure of the dynamical asymmetry encoded in cross-correlations between pairs of observables, which can in turn be used to quantify other physical quantities such as the degree of directed information flow~\cite{10BioPhys_EventOrdering,13NeuroImage_Connectivity,23ohgaprl} and circulations in the space of observables~\cite{76PTEP_IrreversibleCirculation,18Rep_BrokenDetailedBalance,19PRE_ExperimentalMetric}. Equation~\eqref{finite2} thus extends thermodynamic bounds on the asymmetry of cross-correlation functions \cite{23ohgaprl,Shiraishi23Asym,liang2023pre,Tan24asym,Gu24Asym,Aslyamov25arXiv}, linking experimentally accessible measurements to fundamental irreversibility. 

{In Sec. III of \cite{supp_mat}, we use a four-state push-pull model \cite{23ohgaprl} with FG states labeled by $(a,b)\in\{0,1\}^2$ to provide a transparent illustration of both Eq.~\eqref{CG-EPR-corr} and the looser finite-time bound Eq.~\eqref{finite2}. There, we extract information about irreversibility from coarse information: we choose the observables ${O'}^{(a,b)}_{1}=x(1-b)$, ${O'}^{(a,b)}_{2}=(1-x)a$, ${O'}^{(a,b)}_{3}=xb$, and ${O'}^{(a,b)}_{4}=(1-x)(1-a)$ with a parameter $x \in [0,1]$, while the FG states $(a,b)$ are not accessible. 
In the limits $x=0$ and $x=1$, this construction reduces to deterministic two-state coarse-graining where no EPR can be inferred.} 



\textit{Measuring irreversibility by density cross-correlation functions.}---{To demonstrate the second interpretation in a realistic setup,} we consider an interacting many-particle system undergoing diffusion under external forces and interparticle interactions. The FG state $i$ corresponds to the positions of all particles, which form a high-dimensional continuous vector. We implicitly consider a continuous limit of our general framework.


We artificially divide the real space into virtual boxes labeled by $m$, and then map each FG state into the particle number density distribution in these boxes, i.e., choosing $T (m|i)= \rho_m^i \equiv n_m^i/n_{\rm tot}$, where $\rho^i_m$ is the number density in region $m$ at the FG state $i$, $n_m^i$ is the particle number in region $m$ of state $i$, and $n_{\rm tot}=\sum_m n_m^i$ is the total particle number; see Fig.~\ref{1}(b) for illustration. If we assume the conservation of the particle number in the system of interest, $n_{\rm tot}$ will be independent of $i$. Then, the probability of finding the system in the CG state $m$ is given by $p^{\rm cg}_m(t)=\sum_i\, \rho_m^i p_i(t) \eqqcolon \langle \rho_m(t)\rangle$. {Crucially, this is already a random coarse-graining rather than a deterministic lumping: a single microscopic configuration $i$ is mapped to the whole probability vector $\{\rho_m^i\}_m$, so one configuration generally contributes to several coarse-grained outcomes simultaneously.}
More generally, different particle species could have different weights in the mapping. This procedure is inspired by an example proposed in \cite{bao2023universal} by one of the authors, which was not explicitly explained there.


In this case, the CG EPR can be rewritten with the cross-correlation function between the number density in the virtual box $m$ and $n$,
\begin{equation}\label{density_corr}
    C_{nm}^{t,\tau} \equiv \langle \rho_n(t) \rho_m (t+\tau) \rangle \coloneqq 
    \sum_{i,j} \,\rho_n^i\rho_m^j \mathcal{P}[j,t+\tau;i,t].
\end{equation}
Substituting Eq.~\eqref{density_corr} into Eq.~\eqref{finite2}, we obtain an operational estimator of the EPR {$\sigma_{[t,t+\tau]}^{\rm est}$} solely using the density correlation function. 

{Then, each term in the estimator $\sigma_{[t,t+\tau]}^{\rm est}$ can be interpreted as} the irreversibility within the combined region of boxes $m$ and $n$, which permits the inference of irreversibility when only part of the system is accessible \footnote{In steady states or periodic steady states with the period being $\tau$ and the protocol being time-symmetric, the symmetrized pair contributions in Eq.~\eqref{finite2} provide nonnegative local measures of irreversibility. For general time-dependent protocols, the nonnegative local decomposition is obtained using the rewritten form given in Sec. I of the SM.}. Rather than merely providing a single global metric, measuring {individual} terms reveals the spatial patterns of irreversibility throughout the system.

Our proposed protocol for obtaining lower bounds is highly experimentally feasible, as it only involves the counting of particle numbers within virtual boxes, which has been implemented in a recent paper \cite{count24prx} to evaluate diffusion coefficients in many-body systems. Indeed, particle number measurement is a ubiquitous tool in experimental studies of many-body systems, ranging from cold atoms~\cite{gross2017quantum} to colloidal suspensions and active matter systems~\cite{palacci2013living}. This universality ensures that our framework is directly compatible with standard experimental observables, facilitating practical application across diverse platforms. A key operational advantage of our protocol is that it does not require precise tracking of particle positions or knowledge of particle species, marking a significant improvement over conventional approaches. 

\begin{figure}
    \centering
    \includegraphics[width=0.8\linewidth]{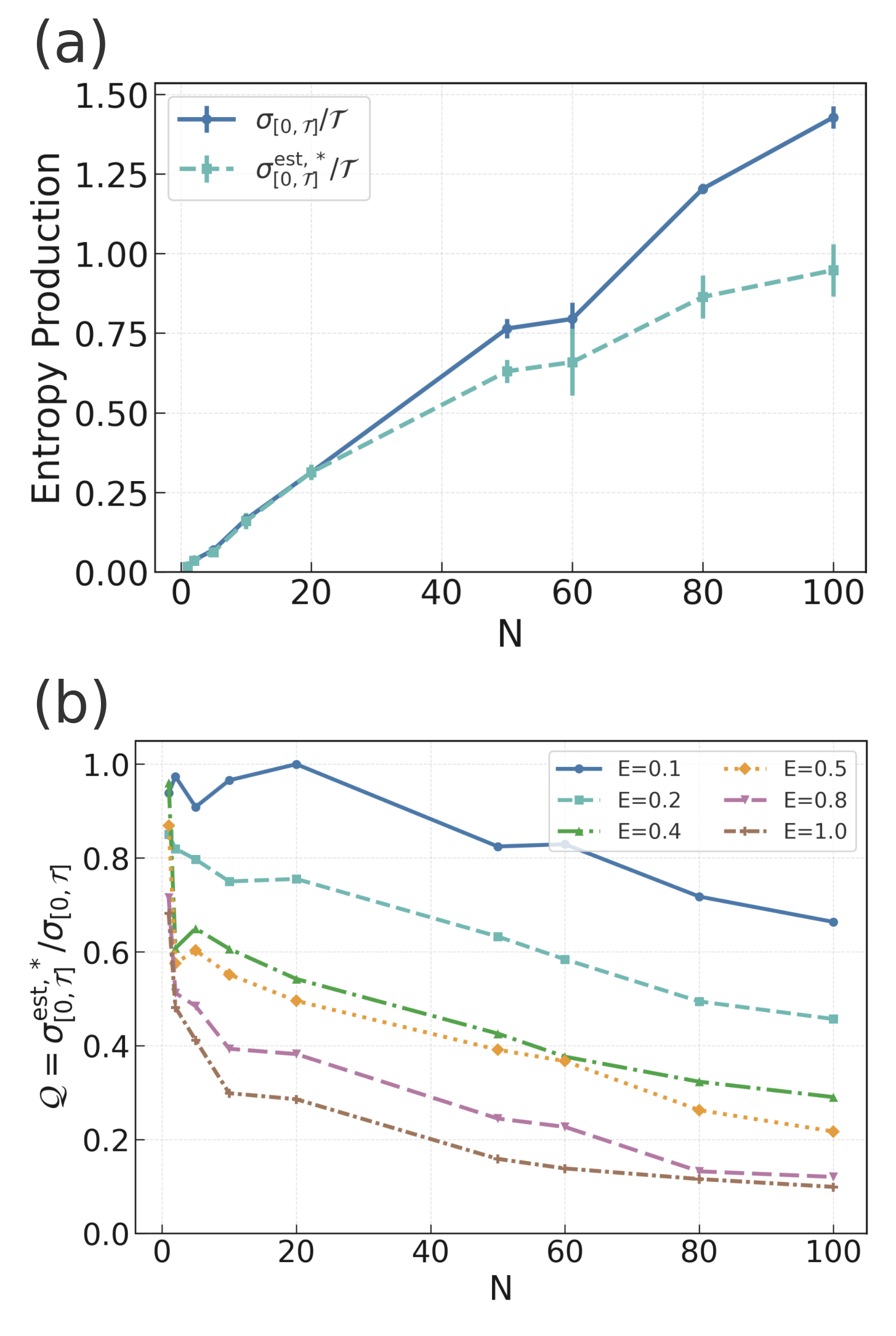}
    \caption{{Estimation of the average EPR over one driving period in the periodically driven many-body example. The period is $1.0.$ (a) True average EPR and optimized estimator versus $N$ for $E_0=E_1=0.1$. (b) Quality factor $\mathcal{Q}:=\sigma^{\rm est,*}_{[0,\mathcal{T}]}/\sigma_{[0,\mathcal{T}]}$ versus $N$ for $E_0=E_1=E$. Error bars indicate the standard deviation over five independent simulations.}}
    \label{2}
\end{figure}

{As a practical application, we consider a periodically driven two-dimensional overdamped system of interacting Brownian particles with heterogeneous radii, charges, and diffusion coefficients to model realistic biological systems. The EP is estimated from the time series of particle numbers in virtual boxes using Eq.~\eqref{finite2}, with the lag time optimized over integer multiples of the driving period. All model definitions, parameters, and simulation procedures are given in the SM. Figure~\ref{2} shows the period-averaged EPR together with the optimized estimator for different particle numbers and driving strengths. Despite using only coarse counting data, the estimator captures a significant fraction of the true EP. {Most conventional bounds, including the TUR, do not apply directly there due to the periodic driving. To benchmark our method, the SM also reports a comparison with a TUR-bound built from the same box-counting observables for a nonequilibrium steady state.
See End Matter for more discussion and possible enhancement of the estimators for many-body settings as in this example.}}

\textit{Discussion and Outlook}---{We introduce a random coarse-graining framework that extends conventional deterministic lumping to probabilistic mappings. This framework provides a natural description of imprecise measurements and, more generally, converts the asymmetry of cross-correlations between suitably chosen observables into experimentally accessible measures of irreversibility. In the many-body implementation, these measures rely solely on cross-correlation functions of particle-number density between virtual boxes, requiring no trajectory data or model details. The framework is experimentally feasible and applicable to a wide range of stationary or non-stationary systems, including interacting many-body systems.} It provides a systematic approach to study irreversibility in complex systems. Notably, meaningful inferences can be drawn by accessing only a subset of the system---common in real-world applications. Additionally, our approach naturally reveals the spatial distribution of irreversibility, offering richer insights than a global single-value measure. 

{Our work raises numerous open questions and suggests promising applications.}  
For instance, it could help estimate relaxation timescales in complex systems. Investigating memory effects within coarse-grained dynamics is another intriguing direction. Studies of information thermodynamics and response theory \cite{Basu18CGresponse} in our CG dynamics are also compelling future directions. Given its flexibility and broad applicability, our framework may find applications in other fields, offering a versatile tool for analyzing diverse complex systems. 

While finalizing this work, we became aware of a study on the thermodynamics of faulty coarse-graining that recently appeared~\cite{Jann25faulty}, which addresses the random coarse-graining of trajectories. 


\begin{acknowledgments}  
R.B. is supported by JSPS KAKENHI Grant No.\ 25KJ0766. N.O. is supported by JSPS KAKENHI Grant No.\ 23KJ0732. S.I. is supported by JSPS KAKENHI Grants No.\ 22H01141, No.\ 23H00467, and No.\ 24H00834, JST ERATO Grant No.\ JPMJER2302, and UTEC-UTokyo FSI Research Grant Program. 
\end{acknowledgments}

\bibliographystyle{unsrt}
\bibliography{bibfile}
\section*{End matter}

\textit{Derivation of the effective master equation \eqref{EME}}:
Taking the time derivative of the definition of the CG states $p^{\rm cg}_{m}(t)=\sum_i T (m|i) p_i(t)$ gives
\begin{align}
    & \frac{d}{dt} p^{\rm cg}_{m}(t) 
    \nonumber \\
    &= \sum_i T (m|i) \sum_{j(\neq i)}[ k_{ij}(t) p_j(t) - k_{ji}(t) p_i(t)]
    \nonumber \\
    & =\sum_{n} \sum_{i, j|i \neq j} T (m|i)T (n|j) [k_{ij}(t) p_j(t) -k_{ji}(t) p_i(t)]
    \nonumber \\
    & =\sum_{n (\neq m)} \sum_{i, j|i \neq j} [T (m|i)T (n|j) - T (n|i)T (m|j)] k_{ij}(t) p_j(t),
\end{align}
where we inserted $\sum_n T (n|j)=1$ in the second equality. This allows us to identify the effective transition rates
\begin{equation}
    k^{\rm cg}_{mn}(t)\coloneqq \sum_\nu \sum_{i,j|i \neq j} T (m|i) k^\nu_{ij}(t)T (n|j)\frac{p_j(t)}{p^{\rm cg}_n(t)} 
\end{equation}
for $n\neq m$, where we inserted $k_{ij}(t)=\sum_\nu k_{ij}^\nu (t)$.
We also identify the CG probability fluxes as $k^{\rm cg}_{mn}(t) p^{\rm cg}_n(t)=\sum_{i,j|i\neq j} T (m|i)T (n|j) k_{ij}(t) p_j(t)$.

\textit{The coarse-grained EPR:}
Here we provide details of the decomposition of the true EPR into three terms including the CG EPR, i.e., $\dot{\sigma}(t) = \dot{\sigma}^{\rm cg}(t) + \dot{\sigma}^{\text{inn}}(t)+\dot{\sigma}^{\text{tran}}(t)$. We omit the time dependence whenever obvious. We start with the EPR in the CG level and use the log-sum inequality to get
\begin{align}
    \dot{\sigma}^{\rm cg} 
    & =\sum_{m,n|m\neq n} k^{\rm cg}_{mn} p^{\rm cg}_n \ln{\frac{k^{\rm cg}_{mn} p^{\rm cg}_n}{k^{\rm cg}_{nm} p^{\rm cg}_m}}
    \notag \\
    & = \sum_{m,n|m\neq n} \Biggl\{\left[\sum_\nu \sum_{i,j|i\neq j} T (m|i)T (n|j) k^\nu_{ij} p_j\right] 
    \notag \\
    & \qquad \qquad \times \ln{\frac{\sum_\nu \sum_{i,j|i\neq j} T (m|i)T (n|j) k^\nu_{ij} p_j}{\sum_\nu \sum_{i,j|i\neq j} T (m|i)T (n|j) k^\nu_{ji} p_i}} \Biggr\}
    \notag \\
    & \leq \! \! \! \! \! \! \sum_{m,n,\nu, i,j|i\neq j, m\neq n} \! \! \! \!\! \!  T (m|i)T (n|j) k^\nu_{ij} p_j \ln{\frac{ T (m|i)T (n|j) k^\nu_{ij} p_j}{ T (m|i)T (n|j) k^\nu_{ji} p_i}} 
    \notag \\
    & = \sum_{\nu, i,j|i\neq j} \sum_{m,n} T (m|i)T (n|j) k^\nu_{ij} p_j \ln{\frac{k^\nu_{ij} p_j}{k^\nu_{ji} p_i}} 
    \notag \\
    & \quad - \sum_m \sum_{\nu, i,j|i\neq j} T (m|i)T (m|j) k^\nu_{ij} p_j \ln{\frac{k^\nu_{ij} p_j}{k^\nu_{ji} p_i}} .
    \label{cgepr-proof}
\end{align}
The first term on the last side of Eq.~\eqref{cgepr-proof} is equal to $\dot{\sigma}$, as follows from $\sum_m T (m|i)=\sum_n T (n|j) =1$. We identify the second term as the contribution $-\dot{\sigma}^{\text{inn}}$, i.e.,
\begin{equation}
    \dot{\sigma}^{\text{inn}}\coloneqq \sum_m \sum_\nu \sum_{i,j|i\neq j} T (m|i)T (m|j) k^\nu_{ij} p_j \ln{\frac{k^\nu_{ij} p_j}{k^\nu_{ji} p_i}} \geq 0.
\end{equation}
This contribution is interpreted as the sum of the hidden EPR within each CG state $m$. We thus get 
\begin{equation}
    \dot{\sigma}^{\rm cg} \leq \dot{\sigma} - \dot{\sigma}^{\text{inn}} \leq \dot {\sigma}.
\end{equation}
The remaining part $\dot{\sigma}^{\text{tran}}$ comes from the gap between the two sides of the inequality in Eq.~\eqref{cgepr-proof}. The gap is rearranged as
\begin{align}
    \dot{\sigma}^{\text{tran}}
    &\coloneqq \dot{\sigma} - \dot {\sigma }^{\rm cg} - \dot{\sigma}^{\rm inn} 
    \notag \\
    &= \sum_{m \neq n} k_{mn}^{\rm cg}p_n^{\rm cg} \sum_{\nu, i, j|i\neq j}Q_{mn}(\nu,i,j)\ln{\frac{Q_{mn}(\nu,i,j)}{Q_{nm}(\nu,j,i)}}\geq0,
\end{align}
where $Q_{mn}(\nu,i,j) \coloneqq {T (m|i) T (n|j)  k^\nu_{ij} p_j}/(k_{mn}^{\rm cg}p_n^{\rm cg})$ is a normalized probability distribution over $\{(\nu,i,j)\,|\,i\neq j\}$. This contribution $\dot{\sigma}^{\text{tran}}$ is a weighted sum of the Kullback--Leibler divergence between $Q_{mn}(\nu,i,j)$ and $Q_{nm}(\nu, j,i)$, and therefore nonnegative. Physically, this contribution arises from multiple FG fluxes $k^\nu_{ij}p_j$ comprising the forward/backward CG fluxes $k_{mn}^{\rm cg}p_n^{\rm cg}$ and $k_{nm}^{\rm cg}p_m^{\rm cg}$.
In the deterministic CG, all three contributions reduce to their counterparts derived in \cite{esposito12pre}.

\textit{Measure of irreversibility in terms of cross-correlations:} We start by rewriting the effective directed traffic $k^{\rm cg}_{mn} p^{\rm cg}_n$ with the cross-correlation functions defined in {Eq.~\eqref{corr-def}.} 
Using {the identification $T (m|i)=O_m^i$ and} the expansion
\begin{equation}
    p[j,t+\Delta t;i,t] = \delta_{ji}p_i(t)+k_{ji}(t)p_i(t)\Delta t + O(\Delta t^2),
\end{equation}
it is easy to show that
\begin{equation}
    k^{\rm cg}_{mn}(t) p^{\rm cg}_n(t)  =  \frac{C_{nm}^{t,\Delta t} - B_{nm}^{t,\Delta t}}{\Delta t} +O(\Delta t).
\end{equation}
Inserting this expression into the definition of $\dot{\sigma}^{\rm cg}$ gives an experimentally feasible measure of irreversibility,
\begin{align}
    \dot{\sigma}^{\rm cg} &= \lim_{\Delta t\to 0} \sum_{m,n|m\neq n}\frac{C^{t,\Delta t}_{nm} - B^{t,\Delta t}_{nm}}{\Delta t} \ln \frac{C^{t,\Delta t}_{nm} - B^{t,\Delta t}_{nm}}{C^{t,\Delta t}_{mn} - B^{t,\Delta t}_{mn}}
    \notag \\
    &= \lim_{\Delta t\to 0} \sum_{m,n|m> n}\frac{C^{t,\Delta t}_{nm} - C^{t,\Delta t}_{mn}}{\Delta t} \ln \frac{C^{t,\Delta t}_{nm} - B^{t,\Delta t}_{nm}}{C^{t,\Delta t}_{mn} - B^{t,\Delta t}_{mn}},
\end{align}
where the second equality follows from $B^{t,\Delta t}_{mn} =B^{t,\Delta t}_{nm}$.

Furthermore, we can drop the term $B_{mn}^{t,\Delta t}$ by using $(a-b)\ln [(a-c)/(b-c)]\geq (a-b)\ln (a/b)$ for $a \geq 0,b\geq0,\ a \geq c,b\geq c$: {
\begin{align}
\dot{\sigma}^{\rm cg} &\geq \lim_{\Delta t \to 0} \sum_{m> n}\frac{C^{t,\Delta t}_{nm} - C^{t,\Delta t}_{mn}}{\Delta t} \ln \frac{C^{t,\Delta t}_{nm} }{C^{t,\Delta t}_{mn}}\nonumber \\
    &=\lim_{\Delta t \to 0} \sum_{m, n}\frac{C^{t,\Delta t}_{nm}}{\Delta t} \ln \frac{C^{t,\Delta t}_{nm} }{C^{t,\Delta t}_{mn}}.
\end{align}
Here, $C^{t,\Delta t}_{nm} \geq B^{t,\Delta t}_{nm}$ because of the nonnegativity of $k^{\rm cg}_{mn}(t) p^{\rm cg}_n(t)$.}

\textit{Proof of the finite-time bounds leading to Eq.~\eqref{finite2}---}We first define a two-time averaged EPR: 
\begin{equation}
    \frac{\sigma^{\text{two-time}}_{[t,t+\tau]}}{\tau}\coloneqq \frac{1}{\tau} \sum_{i,j}K^{t,\tau}_{ij}p_j(t)\ln{\frac{K^{t,\tau}_{ij}p_j(t)}{K^{\dagger,t,\tau}_{ji}p_i(t+\tau)}}
    \label{finite1}
\end{equation}
where $K^{t,\tau}_{ij}\coloneqq \mathcal{P}[i,t+\tau|j,t]$ 
is the finite-time propagator (conditional probability) generated by the transition rates $\{k_{ij}(s)\}_{t\leq s \leq t+\tau }$, and $K^{\dagger, t,\tau}_{ij}\coloneqq \mathcal{P}^{\dagger}[i,t+\tau|j,t]$ is the propagator generated by the backward (time-reversed) transition rates $\{k_{ij}(2t+\tau-s)\}_{t\leq s \leq t+\tau}$. $\sigma^{\text{two-time}}_{[t,t+\tau]}$ could be regarded as the EP under both spatial and temporal coarse-graining, and it is a lower bound of the true EP.

Then, we prove that $ \sigma_{[t,t+\tau]}\geq \sigma^{\text{two-time}}_{[t,t+\tau]}. $
Without losing generality, we discuss the interval $[0, \tau]$ instead of the interval $[t, t+\tau]$. For simplicity, we present the finite-time proof in the single-bath case. 
To introduce an expression of $\sigma_{[0,\tau]}$,  we consider a trajectory $\gamma \equiv \{\gamma_{t}\}_{0 \leq  t \leq \tau}$ over $[0,\tau]$, where $\gamma_{t}$ is the FG state at time $t$. The time-reversed trajectory is defined as $\gamma^{\dagger} \coloneqq \{\gamma_{\tau-t}\}_{0 \leq  t \leq \tau}$. The true EP is rewritten as~\cite{seifert05prl,broeck07dissipation}
\begin{equation}
\sigma_{[0,\tau]} = \int d\gamma\,\mathbb{P}(\gamma)\ln\frac{\mathbb{P}(\gamma)}{\mathbb{P}^{\dagger}(\gamma^{\dagger})},
\label{ep-traj}
\end{equation}
where $\mathbb{P}(\gamma)$ is the probability of realizing $\gamma$ under the transition rates $\{k_{ij}(t)\}_{0 \leq  t \leq \tau}$ {with the initial distribution $p_i(0)$}, and $\mathbb{P}^{\dagger}(\gamma)$ is the probability under the time-reversed transition rates $\{k_{ij}(\tau-t)\}_{0 \leq  t \leq \tau}$ {with the initial distribution $p_i(\tau)$}. The processes under the forward protocol $\{k_{ij}(t)\}_{0 \leq  t \leq \tau}$ and the backward protocol $\{k_{ij}(\tau-t)\}_{0 \leq  t \leq \tau}$ are the forward and backward processes mentioned in the main text, respectively.

{We use these path probabilities to rewrite the joint probabilities in the main text, $\mathcal{P}[i,\tau;j,0]$ and $\mathcal{P}^\dagger[i,\tau;j,0]$}.
By introducing the set of trajectories with a specified final and initial states,
\begin{equation}
\Lambda(i,j) \coloneqq \{\gamma\mid\gamma_{\tau} = i\;\text{and}\;\gamma_{0} = j\},
\end{equation}
{the joint probability that the system is in $j$ at time 0 and in $i$ at time $\tau$ under the transition rate $\{k_{ij}(t)\}_{0 \leq  t \leq \tau}$ and initial distribution $p_i(0)$ is $\mathcal{P}[i,\tau;j,0] = \int_{\Lambda(i,j)}\mathbb{P}(\gamma)d\gamma$, and the same probability under the rate $\{k_{ij}(\tau-t)\}_{0 \leq  t \leq \tau}$ and initial distribution $p_i(\tau)$ is $\mathcal{P}^\dagger[i,\tau; j,0] = \int_{\Lambda(i,j)}\mathbb{P}^{\dagger}(\gamma)d\gamma$.}
We introduce the conditional probabilities, $\mathbb{P}(\gamma\vert i,j) \coloneqq \openone_{\Lambda(i,j)}(\gamma)\mathbb{P}(\gamma)/\mathcal{P}[i,\tau;j,0]$ and $\mathbb{P}^{\dagger}(\gamma\vert i,j) \coloneqq \openone_{\Lambda(i,j)}(\gamma)\mathbb{P}^{\dagger}(\gamma)/\mathcal{P}^\dagger[i,\tau;j,0]$, which are the probabilities of the path $\gamma$ conditional to given final and initial states $(i,j)$.

We then rewrite the true EP in Eq.~\eqref{ep-traj} as
\begin{align}
\sigma_{[0,\tau]} &  = \sum_{i,j}\int_{\Lambda(i,j)}\!\! \mathbb{P}(\gamma)\ln\frac{\mathbb{P}(\gamma)}{\mathbb{P}^{\dagger}(\gamma^{\dagger})}d\gamma\nonumber \\
 &  = \sum_{i,j}\int_{\Lambda(i,j)}\!\!\mathcal{P}[i,\tau;j,0]\mathbb{P}(\gamma\vert i,j)\ln\frac{\mathbb{P}(\gamma\vert i,j)}{\mathbb{P}^{\dagger}(\gamma^{\dagger}\vert j,i)}d\gamma
 \nonumber \\
 & \quad  +\sum_{i,j}\int_{\Lambda(i,j)}\!\! \mathcal{P}[i,\tau;j,0] \mathbb{P}(\gamma\vert i,j)\ln\frac{\mathcal{P}[i,\tau;j,0]}{\mathcal{P}^\dagger[j,\tau;i,0]}d\gamma,
\label{ep-conditionalKL}
\end{align}
where we used the fact that $\gamma \in \Lambda(i,j)$ if and only if $\gamma^{\dagger} \in \Lambda(j,i)$. The first term of Eq.~\eqref{ep-conditionalKL} is a conditional Kullback--Leibler divergence, and hence it is nonnegative. The second term of Eq.~\eqref{ep-conditionalKL} can be simplified using $\int_{\Lambda(i,j)}\mathbb{P}(\gamma\vert i,j)d\gamma = 1$. By dropping the first term, we get a lower bound,
\begin{equation}
\sigma_{[0,\tau]} \geq \sum_{i,j}\mathcal{P}[i,\tau;j,0]\ln\frac{\mathcal{P}[i,\tau;j,0]}{\mathcal{P}^\dagger[j,\tau;i,0]} = \sigma^{\text{two-time}}_{[0,\tau]},
\end{equation}
where the second equality follows from $\mathcal{P}[i,\tau;j,0]=K^{0,\tau}_{ij}p_j(0)$ and $\mathcal{P}^\dagger[i,\tau;j,0]=K^{\dagger,0,\tau}_{ij}p_j(\tau)$. Notably, a similar result has been discussed for time-independent transition rates in \cite{Tan24asym}.

 We next prove the inequality $\sigma^{\text{two-time}}_{[t,t+\tau]} \geq \sigma_{[0,\tau]}^{\rm est}$ [Eq.~\eqref{finite2}]. By replacing $k_{ij}\Delta t$ and $k_{ji}\Delta t$ in the original derivation of the short-time lower bound [cf.~Eq.~\eqref{cgepr-proof}] with the finite-time propagators $K_{ij}^{0,\tau}$ and $K_{ji}^{\dagger,0,\tau}$, respectively, we obtain a lower bound on $\sigma^{\text{two-time}}_{[0,\tau]}$ as follows,
\begin{align}
    &\sigma^{\text{two-time}}_{[0,\tau]} \notag \\
    & = \! \!\sum_{m,n, i,j}\! \!  T (m|i)T (n|j) K_{ij}^{0,\tau} p_j (0)\ln{\frac{ T (m|i)T (n|j) K_{ij}^{0,\tau} p_j(0)}{ T (m|i)T (n|j) K_{ji}^{\dagger,0,\tau} p_i(\tau)}} \notag \\
    & \geq \sum_{m,n} \Biggl\{\left[\sum_{i,j} T (m|i)T (n|j) K_{ij}^{0,\tau} p_j (0)\right] 
    \notag \\
    & \qquad \qquad \times \ln{\frac{ \sum_{i,j} T (m|i)T (n|j) K_{ij}^{0,\tau} p_j (0)}{ \sum_{i,j} T (m|i)T (n|j)K_{ji}^{\dagger,0,\tau} p_i(\tau)}} \Biggr\}
    \label{two_time_derivation}
\end{align}
Applying $T (m|i) = O_m^i$ to Eq.~\eqref{two_time_derivation} yields the inequality in Eq. \eqref{finite2}:
\begin{equation}
   \sigma^{\text{two-time}}_{[0,\tau]}\geq\sum_{m,n}C^{0,\tau}_{nm}  \ln \frac{C^{0,\tau}_{nm} }{C^{\dagger,0,\tau}_{mn}}= \sigma_{[0,\tau]}^{\rm est}.
\end{equation} 
{
\textit{Possible enhancement of the counting estimator---}In the weak-interaction limit, the $N$-scaling of this density-counting estimator can be derived explicitly and yields an enhanced lower bound with an $N$-dependent prefactor, while for general interacting systems no universal $N$-only enhancement is possible because the lower bound can scale linearly with $N$ in specific cases; see Sec.~VI of the SM~\cite{supp_mat}.

\textit{More discussion on the illustrative example---}
It is important to emphasize that coarse-graining is not expected to preserve the dependence of the true EP on parameters such as particle number or driving amplitude: as the dynamics changes, the fraction of dissipative cycles visible through a fixed coarse-graining can vary. Sublinear or otherwise nontrivial scaling of the estimator is therefore a generic consequence of information loss, not a pathology unique to random coarse-graining. 

Additionally, we explain here why the estimator performs well when $N$ is small. In this case, our intuition is that fewer dissipative cycles are hidden \cite{blom_milestoning_2024} by the fixed coarse-graining, so a larger fraction of the dissipation remains visible at the coarse level. This makes the
bound comparatively tighter. 
}

\end{document}


\title{Supplemental Material}
\author{Ruicheng Bao}
\thanks{Corresponding author: ruicheng@g.ecc.u-tokyo.ac.jp}
\affiliation{Department of Physics, Graduate School of Science, The University of Tokyo, 7-3-1 Hongo, Bunkyo-ku, Tokyo 113-0033, Japan}
\author{Naruo Ohga}
\affiliation{Department of Physics, Graduate School of Science, The University of Tokyo, 7-3-1 Hongo, Bunkyo-ku, Tokyo 113-0033, Japan}
\author{Sosuke Ito}
\affiliation{Department of Physics, Graduate School of Science, The University of Tokyo, 7-3-1 Hongo, Bunkyo-ku, Tokyo 113-0033, Japan}
\affiliation{Universal Biology Institute, Graduate School of Science, The University of Tokyo, 7-3-1 Hongo, Bunkyo-ku, Tokyo 113-0033, Japan}

\maketitle

\tableofcontents

\section{Decomposition of the estimator into local contributions for general driven dynamics}
We assume  $K_{ji}^{\dagger,t,t+\tau}=K_{ji}^{t,t+\tau}$ and $p_i(t+\tau)=p_i(t)$, implying $C^{\dagger, t,\tau}_{mn}=C^{t,\tau}_{mn}$. In this case, the estimator $\sigma_{[t,t+\tau]}^{\rm est}$ can be rewritten as
 \begin{equation}
     \sigma_{[t,t+\tau]}^{\rm est}=\sum_{m,n|m>n}(C^{t,\tau}_{nm}-C^{ t,\tau}_{mn}) \ln{\frac{C^{t,\tau}_{nm}}{C^{ t,\tau}_{mn}}} 
 \end{equation}
where each term in the sum provides a \textit{nonnegative measure} of irreversibility. For more general situations with arbitrary time-dependent protocols, $\sigma_{[t,t+\tau]}^{\rm est}$ can still be rewritten as a sum of nonnegative measures of spatially local irreversibility:
\begin{equation}
 \sigma_{[t,t+\tau]}^{\rm est} = \sum_{m,n} \left[ C^{t,\tau}_{nm} \ln \left( \frac{C^{t,\tau}_{nm}}{C^{\dagger, t,\tau}_{mn}} \right) -C_{nm}^{t,\tau}+C_{mn}^{\dagger,t,\tau} \right],
\end{equation}
where we use $\sum_{nm} C^{t,\tau}_{nm} = \sum_{nm} C^{\dagger,t,\tau}_{mn} =1$. Each term in this sum is nonnegative. This is shown by using $f(t)=t\ln t - t +1\geq 0$ for nonnegative $t$ and rewriting the term as $C_{mn}^{\dagger,t,\tau} f(C_{nm}^{t,\tau} /C_{mn}^{\dagger, t,\tau} )$.

\section{Imprecise measurements: one-parameter channels, exact recovery, and monotonicity}

This section supplements the brief discussion of imprecise measurements in the main text. We first record the simplest one-parameter measurement-error channel on a uniform four-state ring. We then prove a general exact-recovery criterion for interior error probabilities and state a monotonicity theorem for pairwise adjacent-confusion channels.

\subsection{Direct one-parameter channel on the uniform four-state ring}

Consider the uniform four-state ring $1\to2\to3\to4\to1$ with clockwise and counterclockwise rates $k_{21}=k_{32}=k_{43}=k_{14}\equiv k_+=e^{A/2}$ and $k_{12}=k_{23}=k_{34}=k_{41}\equiv k_-=e^{-A/2}$. The steady state is uniform, $p_i^{\rm ss}=1/4$, and the true steady-state entropy production rate is
\begin{equation}
\dot{\sigma}=(k_+-k_-)\ln\frac{k_+}{k_-}=2A\sinh\left(\frac{A}{2}\right).
\end{equation}
We consider the direct one-parameter nearest-neighbor measurement-error channel  $T(m|i)=T(m|i; \epsilon)$, defined as
\begin{equation}
T(m|i;\varepsilon)=(1-\varepsilon)\delta_{m,i}+\varepsilon\,\delta_{m,i-1},
\qquad 0\le \varepsilon\le 1,
\end{equation}
with indices understood modulo $4$. The observed clockwise nearest-neighbor traffic is multiplied by $(1-\varepsilon)^2+\varepsilon^2$, and the same prefactor multiplies the observed counterclockwise traffic. The remaining contributions are time-symmetric and do not contribute to entropy production. Hence
\begin{equation}
\dot{\sigma}^{\rm cg}(\varepsilon)=\big[(1-\varepsilon)^2+\varepsilon^2\big]\dot{\sigma}.
\label{eq:uniform-ring-formula}
\end{equation}
Here, the argument $\varepsilon$ in $\dot{\sigma}^{\rm cg}(\varepsilon)$ explicitly indicates the dependence on $\varepsilon$, and we will use this notation in the supplementary material going forward. Therefore, $\dot{\sigma}^{\rm cg}(\varepsilon)$ decreases monotonically on $[0,1/2]$, attains its minimum
\begin{equation}
\dot{\sigma}^{\rm cg}(1/2)=\frac12\dot{\sigma}.
\end{equation}
For $\varepsilon\geq 1/2$, $\dot{\sigma}^{\rm cg}(\varepsilon)$ instead increases monotonically
and returns to the true value by deterministic relabeling when $\varepsilon\to 1$.

\subsection{Exact-recovery criterion for interior error probabilities}

We now consider a finite continuous-time Markov jump process in a nonequilibrium steady state. Let $k_{ij}$ be the transition rate from $j$ to $i$, let $p_j$ be the steady-state distribution, and define the steady directed traffic
\begin{equation}
F_{ij}:=k_{ij}p_j.
\end{equation}
The true steady-state entropy production rate is
\begin{equation}
\dot{\sigma}=\sum_{i,j|i\neq j} F_{ij}\ln\frac{F_{ij}}{F_{ji}}.
\label{eq:trueEP_SM}
\end{equation}
After passing both the initial and final microscopic states through the noisy readout channel $T(m|i)=T(m|i; \epsilon)$, defined as
\begin{equation}
T(m|i;\varepsilon)=(1-\varepsilon)\delta_{m,i}+\varepsilon\,\delta_{m,P(i)},
\qquad 0<\varepsilon<1,
\label{eq:general-channel-sm}
\end{equation}
where $P$ is a bijection (permutation) that maps each state to an adjacent state, the observed directed traffic becomes
\begin{equation}
\Phi_{mn}^{(\varepsilon)}:=\sum_{i,j}T(m|i;\varepsilon)\,T(n|j;\varepsilon)\,F_{ij},
\label{eq:observedflux-sm}
\end{equation}
and the coarse-grained entropy production rate is
\begin{equation}
\dot{\sigma}^{\rm cg}(\varepsilon)=\sum_{m,n|m\neq n}\Phi_{mn}^{(\varepsilon)}
\ln\frac{\Phi_{mn}^{(\varepsilon)}}{\Phi_{nm}^{(\varepsilon)}}.
\label{eq:cgEP-sm}
\end{equation}
For each observed pair $(m,n)$, define the confusion cell
\begin{equation}
\mathcal{C}_{mn}:=\Bigl\{(i,j)\,\Big|\,F_{ij}>0,\;
i\in\{m,P^{-1}(m)\},\;
j\in\{n,P^{-1}(n)\}\Bigr\}.
\label{eq:cell-sm}
\end{equation}
For interior $\varepsilon$, the support of each cell is independent of $\varepsilon$; only the weights change.

\begin{theorem}[Cell-affinity criterion]
Assume $F_{ij}>0$ implies $F_{ji}>0$. For any interior error probability $0<\varepsilon<1$, exact recovery
\begin{equation}
\dot{\sigma}^{\rm cg}(\varepsilon)=\dot{\sigma}
\end{equation}
holds if and only if, for every observed pair $(m,n)$, the microscopic affinity
\begin{equation}
A_{ij}:=\ln\frac{F_{ij}}{F_{ji}}
\end{equation}
is constant on the whole cell $\mathcal{C}_{mn}$.
\end{theorem}

\paragraph{Proof.}
Fix an observed pair $(m,n)$ and write
\begin{equation}
a_{ij}=T(m|i;\varepsilon)\,T(n|j;\varepsilon)\,F_{ij},
\qquad
b_{ij}=T(m|i;\varepsilon)\,T(n|j;\varepsilon)\,F_{ji}.
\end{equation}
Then $\Phi_{mn}^{(\varepsilon)}=\sum_{(i,j)\in\mathcal{C}_{mn}} a_{ij}$ and $\Phi_{nm}^{(\varepsilon)}=\sum_{(i,j)\in\mathcal{C}_{mn}} b_{ij}$. Applying the log-sum inequality,
\begin{equation}
\Phi_{mn}^{(\varepsilon)}
\ln\frac{\Phi_{mn}^{(\varepsilon)}}{\Phi_{nm}^{(\varepsilon)}}
\le
\sum_{(i,j)\in\mathcal{C}_{mn}}
a_{ij}\ln\frac{a_{ij}}{b_{ij}}
=
\sum_{(i,j)\in\mathcal{C}_{mn}}
T(m|i;\varepsilon)\,T(n|j;\varepsilon)\,F_{ij}\ln\frac{F_{ij}}{F_{ji}}.
\end{equation}
Summing over $(m,n)$ gives $\dot{\sigma}_{\rm cg}(\varepsilon)\le \dot{\sigma}$. Equality of the log-sum inequality holds exactly when $a_{ij}/b_{ij}=F_{ij}/F_{ji}$ is constant on the whole cell, which is precisely the stated condition. \hfill$\square$

\begin{corollary}[No threshold and no isolated interior point]
Under the same assumptions, if exact recovery holds for one interior value $\varepsilon_*\in(0,1)$, then it holds for every interior value $\varepsilon\in(0,1)$. Therefore there is no phenomenon in which exact recovery starts only above a threshold $\varepsilon_c\in(1/2,1)$, and there is no isolated exact-recovery point $\varepsilon_*\in(1/2,1)$.
\end{corollary}

The criterion depends only on the cell supports \eqref{eq:cell-sm}, and these supports are the same for all $0<\varepsilon<1$. Hence exact recovery away from the endpoints is either an all-interior plateau or it does not occur at all.

\begin{figure}
    \centering
    \includegraphics[width=1.0\linewidth]{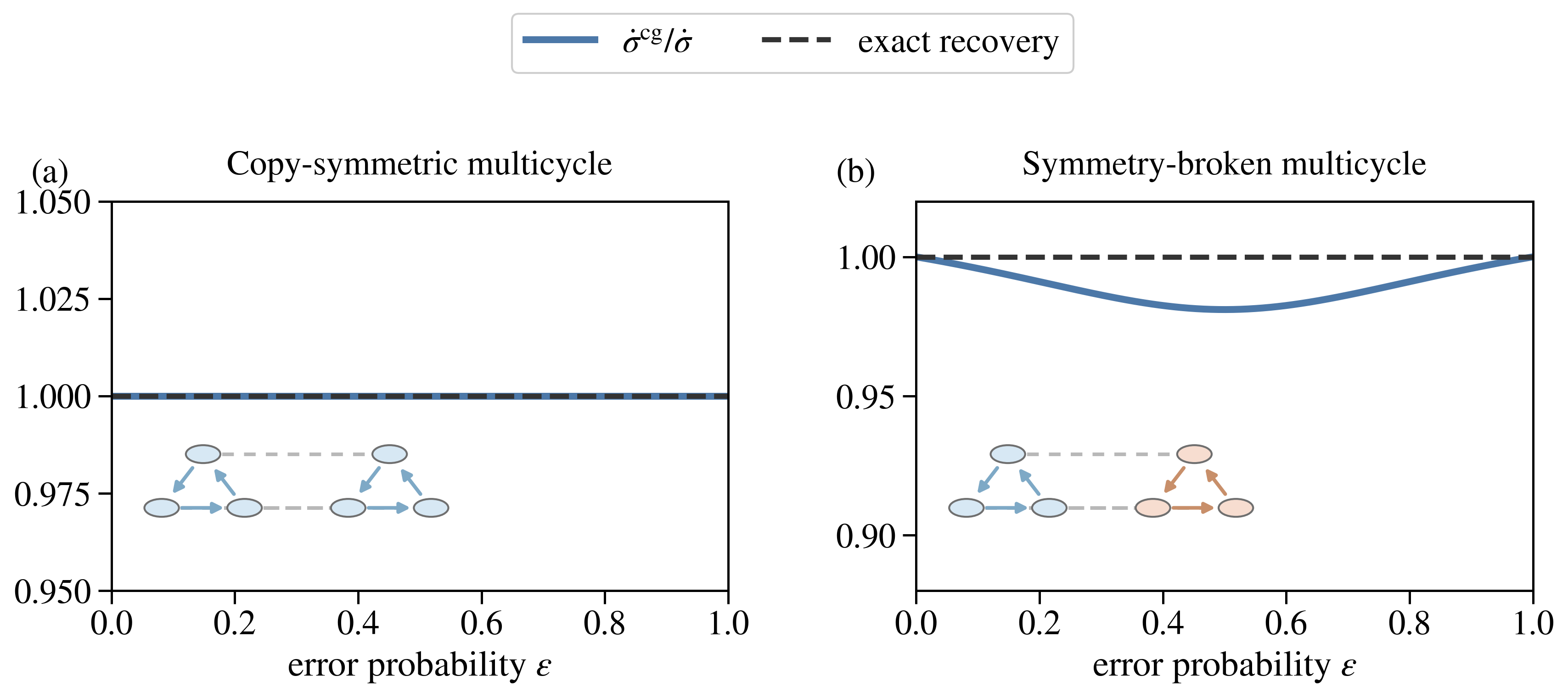}
    \caption{(a) A copy-symmetric nonuniform multicycle exhibits exact recovery for all interior $\varepsilon$. (b) A small symmetry breaking removes that plateau immediately.}
    \label{exact_recovery}
\end{figure}

A broad exact-recovery family follows immediately. Take two copies, $A$ and $B$, of an arbitrary base network, with identical rates on corresponding edges. Connect each pair of corresponding states $i_A$ and $i_B$ by a detailed-balanced bridge $i_A\leftrightarrow i_B$ with equal forward and backward rate. Let the confusing map swap each corresponding pair. The generator then commutes with the swap, the steady state satisfies $p_{i_A}=p_{i_B}$, and each confusion cell merges only (i) two corresponding copy-edges carrying the same affinity or (ii) one bridge pair with zero affinity. Therefore the cell-affinity criterion is satisfied on every cell, yielding $\dot{\sigma}^{\rm cg}(\varepsilon)=\dot{\sigma}$ for all interior $\varepsilon$.
An explicit six-state example for the exact recovery is presented in Fig. \ref{exact_recovery}.

\subsection{Monotonicity for pairwise adjacent-confusion channels}

We finally state the monotonicity theorem referred to in the main text for the simplified class of pairwise adjacent-confusion channels $T(m|i)=(\mathsf{T}(\varepsilon) )_{mi}$, defined as
\begin{equation}
\mathsf{T}(\varepsilon)=(1-\varepsilon)\mathsf{I}+\varepsilon \mathsf{P},
\qquad \varepsilon\in[0,1],
\label{eq:pair-channel-sm}
\end{equation}
where $\mathsf{I}$ is the identity matrix, and $\mathsf{P}$ is the matrix which swaps disjoint adjacent pairs of states and satisfies $\mathsf{P}^2=\mathsf{I}$.

In what follows, we will use the observed traffic in Eq.~\eqref{eq:observedflux-sm} to study the monotonicity of the inferred entropy production with respect to the error probability. For notation's brevity, we define the matrix of the observed traffic as
\begin{equation}\mathsf{F}^{(\varepsilon)}:=\mathsf{T}(\varepsilon) \mathsf{F}[\mathsf{T}(\varepsilon)]^{\top},
\end{equation}
where the original traffic matrix $ \mathsf{F}$ and the observed traffic matrix are defined as $( \mathsf{F} )_{ij}=F_{ij}$ and $( \mathsf{F}^{(\varepsilon)} )_{ij}=\Phi^{(\varepsilon)}_{ij}$. This is simply the Eq.~\eqref{eq:observedflux-sm} in matrix notation. Recall that the observed (coarse-grained) entropy production is given by $\dot{\sigma}_{\rm cg}(\varepsilon)=\Sigma[\mathsf{F}^{(\varepsilon)}]$, where the function $\Sigma[\mathsf{F}]$ is defined as
\begin{equation}
\Sigma[\mathsf{F}]:=\sum_{m,n|m\neq n}F_{mn}\ln\frac{F_{mn}}{F_{nm}}.
\end{equation}

\begin{proposition}[Post-processing monotonicity]
For any traffic matrix $\mathsf{F}$ and any column-stochastic matrix $\mathsf{S}$,
\begin{equation}
\Sigma[\mathsf{S}\mathsf{F}\mathsf{S}^{\top}]\le \Sigma[\mathsf{F}].
\end{equation}
\end{proposition}

\paragraph{Proof.}
This is the same log-sum-inequality argument used in the derivation of the coarse-grained EPR bound: one expands each observed traffic $(\mathsf{S}\mathsf{F}\mathsf{S}^{\top})_{mn}$ as a nonnegative sum over microscopic traffic contributions and then applies the log-sum inequality term by term. \hfill$\square$

\begin{proposition}[Factorization on the left half of the interval]
For any $0\le \varepsilon_1\le \varepsilon_2 \le 1/2$ and $\varepsilon_1 \neq 1/2$, one has
\begin{equation}
\mathsf{T}(\varepsilon_2)=\mathsf{T}(\eta)\mathsf{T}(\varepsilon_1),
\qquad
\eta=\frac{\varepsilon_2-\varepsilon_1}{1-2\varepsilon_1}\in[0,1/2].
\label{eq:factorization-sm}
\end{equation}
For $\varepsilon_1=\varepsilon_2=1/2$, 
 $\eta=0$ and $\mathsf{T}(\varepsilon_2)=\mathsf{T}(\eta)\mathsf{T}(\varepsilon_1)$.
\end{proposition}

\paragraph{Proof.}
Using $\mathsf{P}^2=\mathsf{I}$,
\begin{align}
\mathsf{T}(\eta)\mathsf{T}(\varepsilon_1)
&=\big[(1-\eta)\mathsf{I}+\eta \mathsf{P}\big]\big[(1-\varepsilon_1)\mathsf{I}+\varepsilon_1 \mathsf{P}\big] \\
&=\Big((1-\eta)(1-\varepsilon_1)+\eta\varepsilon_1\Big)\mathsf{I}
 +\Big((1-\eta)\varepsilon_1+\eta(1-\varepsilon_1)\Big)\mathsf{P}.
\end{align}
Choosing $\eta$ as in Eq.~\eqref{eq:factorization-sm} makes the coefficient of $\mathsf{P}$ equal to $\varepsilon_2$. For $\varepsilon_1=\varepsilon_2=1/2$, 
$\mathsf{T}(0)=\mathsf{I}$, and $\mathsf{T}(\varepsilon_2)=\mathsf{T}(0)\mathsf{T}(\varepsilon_1)$ is satisfied. \hfill$\square$

\begin{proposition}[Mirror symmetry]
For every $\varepsilon\in[0,1]$,
\begin{equation}
\mathsf{T}(\varepsilon)=\mathsf{P}\,\mathsf{T}(1-\varepsilon).
\label{mirror1}
\end{equation}
Therefore
\begin{equation}
\dot{\sigma}^{\rm cg}(\varepsilon)=\dot{\sigma}^{\rm cg}(1-\varepsilon),
\qquad
\dot{\sigma}^{\rm err}(\varepsilon)=\dot{\sigma}^{\rm err}(1-\varepsilon),
\end{equation}
where $\dot{\sigma}^{\rm err}(\varepsilon):=\dot{\sigma}-\dot{\sigma}^{\rm cg}(\varepsilon)$.
\end{proposition}

\paragraph{Proof.}
To prove Eq.~\eqref{mirror1}, we used $\mathsf{P}^2=\mathsf{I}$ in Eq.~\eqref{eq:pair-channel-sm}. Hence, $\mathsf{F}^{(\varepsilon)}= \mathsf{P} \mathsf{F}^{(1-\varepsilon)} \mathsf{P}^{\top}$. Since $\Sigma[\mathsf{F}]$ is invariant under simultaneous relabeling of rows and columns, $\dot{\sigma}^{\rm cg}(\varepsilon)= \Sigma[\mathsf{F}^{(\varepsilon)}]= \Sigma[\mathsf{F}^{(1-\varepsilon)}]
= \dot{\sigma}^{\rm cg}(1-\varepsilon)$,
and therefore $\dot{\sigma}^{\rm err}(\varepsilon)= \dot{\sigma}^{\rm err}(1-\varepsilon)$.
\begin{theorem}[Monotonicity of the hidden entropy production]
For the pairwise adjacent-confusion channel in Eq.~\eqref{eq:pair-channel-sm},
\begin{align}
0\le \varepsilon_1\le \varepsilon_2\le 1/2
&\Longrightarrow
\dot{\sigma}^{\rm err}(\varepsilon_1)\le \dot{\sigma}^{\rm err}(\varepsilon_2),
\\
1/2\le \varepsilon_1\le \varepsilon_2\le 1
&\Longrightarrow
\dot{\sigma}^{\rm err}(\varepsilon_1)\ge \dot{\sigma}^{\rm err}(\varepsilon_2).
\end{align}
Equivalently, $\dot{\sigma}^{\rm cg}(\varepsilon)$ is nonincreasing on $[0,1/2]$ and nondecreasing on $[1/2,1]$.
\end{theorem}

\paragraph{Proof.}
Let $0\le \varepsilon_1\le \varepsilon_2\le1/2$ and write $\mathsf{T}(\varepsilon_2)=\mathsf{T}(\eta)\mathsf{T}(\varepsilon_1)$ as in Eq.~\eqref{eq:factorization-sm}. Then
\begin{equation}
\mathsf{F}^{(\varepsilon_2)}=\mathsf{T}(\eta)\,\mathsf{F}^{(\varepsilon_1)}\,\mathsf{T}(\eta)^{\top}.
\end{equation}
Applying post-processing monotonicity to the traffic matrix $\mathsf{F}^{(\varepsilon_1)}$ and the further channel $\mathsf{T}(\eta)$, that is $\Sigma[\mathsf{T}(\eta)\mathsf{F}^{(\varepsilon_1)}\mathsf{T}(\eta)^{\top}]\le \Sigma[\mathsf{F}^{(\varepsilon_1)}]$, gives
\begin{equation}
\dot{\sigma}^{\rm cg}(\varepsilon_2)\le \dot{\sigma}^{\rm cg}(\varepsilon_1),
\end{equation}
which is equivalent to the left-half monotonicity. The right-half statement then follows from the mirror symmetry. \hfill$\square$

\section{Four-state push-pull illustration}

This section gives the detailed minimal example mentioned in the main text. The FG states are labeled by two binary variables, $(a,b)\in\{0,1\}^2$, namely $1=(0,0)$, $2=(1,0)$, $3=(1,1)$, and $4=(0,1)$. Transitions are allowed only between neighboring states on the cycle $1\leftrightarrow2\leftrightarrow3\leftrightarrow4\leftrightarrow1$.

For the nonequilibrium steady-state example, we choose clockwise and counterclockwise transition rates $k_{21}=k_{32}=k_{43}=k_{14}\equiv k_+=e^{A/2}$ and $k_{12}=k_{23}=k_{34}=k_{41}\equiv k_-=e^{-A/2}$ with $A=1$. For the periodically driven case, we use $k_{\pm}(t)=\exp[\pm A(t)/2]$ with $A(t)=A_0+A_1\sin(2\pi t/\mathcal{T})$, and set $A_0=1$, $A_1=0.8$, and $\mathcal{T}=0.2$.

We compare different estimators. First, we define a four-state random coarse-graining $m\in \{1,2,3,4 \}$ through the normalized observables
\begin{equation}
{O'}_{1}^{(a,b)}=x(1-b),\qquad
{O'}_{2}^{(a,b)}=(1-x)a,\qquad
{O'}_{3}^{(a,b)}=xb,\qquad
{O'}_{4}^{(a,b)}=(1-x)(1-a),
\end{equation}
which satisfy ${O'}_{1}^{(a,b)}+{O'}_{2}^{(a,b)}+{O'}_{3}^{(a,b)}+{O'}_{4}^{(a,b)}= {O'}^{(a,b)}_{\rm tot}=1$ for every microscopic state and therefore define a valid four-outcome random map $O_m^{(a,b)}={O'}_{m}^{(a,b)}/{O'}^{(a,b)}_{\rm tot} ={O'}_{m}^{(a,b)}$. The four-state random-coarse-grained entropy production is
\begin{equation}
\dot{\sigma}^{\rm cg}
=
\sum_{\substack{m\in \{1,2,3,4 \},n\in \{1,2,3,4 \}|m\neq n}}
k^{\rm cg}_{mn}p^{\rm cg}_n
\ln
\frac{k^{\rm cg}_{mn}p^{\rm cg}_n}
{k^{\rm cg}_{nm}p^{\rm cg}_m},
\qquad
T^{}(m|(a,b))=O_m^{(a,b)} .
\end{equation}
The same observables are also used in Eq.~(12) of the main text to construct the observable-based bound,
\begin{equation}
\sigma_{[0,\tau]}^{\rm est}
=
\sum_{m\in \{1,2,3,4 \},n\in \{1,2,3,4 \}}
C_{nm}^{0,\tau}
\ln
\frac{C_{nm}^{0,\tau}}
{C_{mn}^{\dagger,0,\tau}},
\qquad
C_{nm}^{0,\tau}
=
\left\langle O_n(0)O_m(\tau)\right\rangle .
\end{equation}

Second, we define a three-state random coarse-graining for $m\in \{{\rm A}, {\rm B}, {\rm C} \}$ by combining these outcomes as
\begin{equation}
T^{}({\rm A}|(a,b))=x(1-b),\qquad
T^{}({\rm B}|(a,b))=xb+(1-x)a,\qquad
T^{}({\rm C}|(a,b))=(1-x)(1-a).
\end{equation}
The three-state random-coarse-grained entropy production is computed by the same expression,
\begin{equation}
\dot{\sigma}^{\rm cg}
=
\sum_{\substack{m\in \{{\rm A}, {\rm B}, {\rm C} \},n\in \{{\rm A}, {\rm B}, {\rm C} \}| m\neq n}}
k^{\rm cg}_{mn}p^{\rm cg}_n
\ln
\frac{k^{\rm cg}_{mn}p^{\rm cg}_n}
{k^{\rm cg}_{nm}p^{\rm cg}_m}.
\end{equation}
At $x=0$ and $x=1$, both constructions reduce to conventional deterministic two-state coarse-graining, so the dissipative four-state cycle becomes invisible and the estimators vanish.

\begin{figure}[t]
    \centering
    \includegraphics[width=\linewidth]{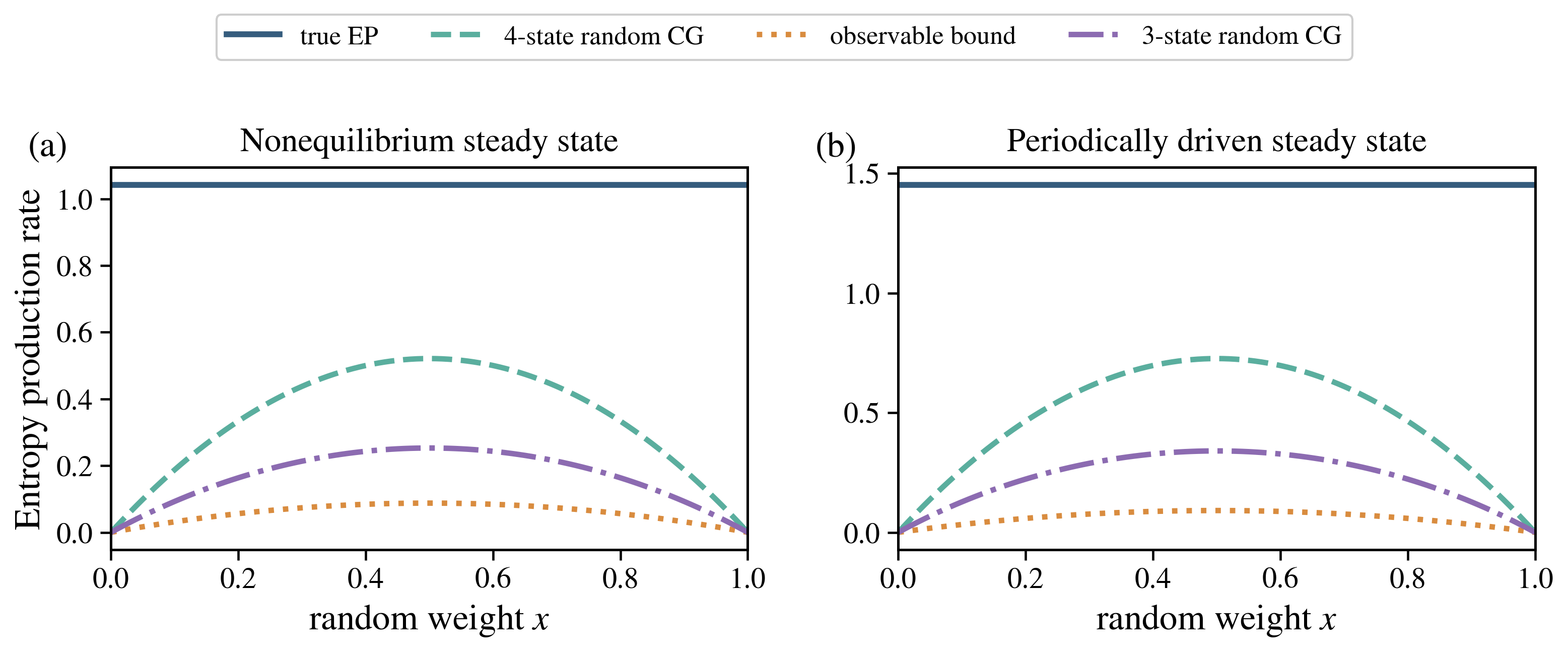}
    \caption{Four-state push-pull illustration of random coarse-graining and the observable-based finite-time bound. Solid dark-blue lines: true entropy production. Dashed green lines: entropy production from the four-state random coarse-graining. Dotted orange lines: observable-based lower bound generated by ${O'}^{(a,b)}_{1}=x(1-b)$, ${O'}^{(a,b)}_{2}=(1-x)a$, ${O'}^{(a,b)}_{3}=xb$, and ${O'}^{(a,b)}_{4}=(1-x)(1-a)$. Dash-dotted purple lines: entropy production from the three-state random coarse-graining. (a) Nonequilibrium steady state with $k_{\pm}=e^{\pm A/2}$ and $A=1$; the observable-based curve shows $\max_{\tau}\sigma_{[0,\tau]}^{\rm est}/\tau$. (b) Periodically driven steady state with $k_{\pm}(t)=\exp\{\pm [A_0+A_1\sin(2\pi t/\mathcal{T})]/2\}$, $A_0=1$, $A_1=0.8$, and $\mathcal{T}=0.2$. The observable-based curve is evaluated over one full period, 
$\sigma^{\rm est}_{[0,\mathcal{T}]}/\mathcal{T}$, without optimizing over the observation time.}
    \label{fig:pushpull-sm}
\end{figure}

Figure~\ref{fig:pushpull-sm} compares the true entropy production, the four-state random-coarse-grained entropy production, the observable-based bound, and the three-state random-coarse-grained entropy production for both the steady and periodically driven cases.

\section{Details of the many-body numerical example}

This section gives the full model definition and numerical protocol for Fig.~3 of the main text, including the interaction potentials, particle heterogeneity, simulation parameters, and the computation of entropy production.

We simulate $N$ interacting Brownian particles in a two-dimensional square domain of size $L = 5$ with periodic boundary conditions. The dynamics of each particle is governed by the overdamped Langevin equation:
\begin{equation}
\dot{\mathbf{x}}_a(t) = \mu_a\mathbf{F}_a(\mathbf{X} (t), t) + \sqrt{2 D_a} \, \boldsymbol{\xi}_a(t),
\end{equation}
where $\mu_a $ is the mobility of the {particle $a$}, {$\mathbf{X} (t) =\{\mathbf{x}_1(t), \dots,  \mathbf{x}_N(t)\}$ is the coordinates of all particles,} $D_a=k_BT\mu_a$ is the diffusion coefficient of particle $a$, and $\boldsymbol{\xi}_a(t)$ is a standard Gaussian white noise satisfying
\begin{equation}
\langle \xi_a^\alpha(t) \xi_b^\beta(t') \rangle = \delta_{ab} \delta_{\alpha\beta} \delta(t - t'),
\end{equation}
{with $\alpha,\beta\in \{1,2\}$ denoting the spatial direction.}

The total force 
\begin{equation}
   \mathbf{F}_a (\mathbf{X} (t) , t)
= -\nabla_{\mathbf{x}_a}\sum_{b|b\ne a} V_{\rm WCA}(r_{ab}(t))
  -\nabla_{\mathbf{x}_a}\sum_{b|b\ne a} V_{\rm spring}(r_{ab}(t))
  + \mathbf{F}^{\rm ext}_a,
\end{equation}
with $r_{ab}(t) = \|{\mathbf{x}}_a (t)-{\mathbf{x}}_b(t) \|$ consists of three parts: a repulsive Weeks–Chandler–Andersen (WCA) interaction,
\begin{equation}
V_{\text{WCA}}(r_{ab}) = 
\begin{cases}
4\epsilon \left[ \left( \frac{\sigma_{ab}}{r_{ab}} \right)^{12} - \left( \frac{\sigma_{ab}}{r_{ab}} \right)^6 \right] + \epsilon, & r_{ab} < 2^{1/6} \sigma_{ab}, \\
0, & \text{otherwise},
\end{cases}
\end{equation}
a harmonic spring potential acting on all particle pairs,
\begin{equation}
V_{\text{spring}}(r_{ab}) = \frac{1}{2} k (r_{ab} - r_0)^2, \quad \text{for } r_{ab} < l_0,
\end{equation}
and a time-dependent external electric force $\mathbf{F}_a^{\text{ext}} = q_a E(t) \hat{x}$ on the $x$ direction 
where
\begin{equation}\label{protocol_periodic}
E(t) = E_0 + E_1 \sin\left( \frac{2\pi t}{\mathcal{T}} \right).
\end{equation}
Here, $\hat{x}$ is the unit vector for the $x$ direction. 
We set $\epsilon = 1$, $k = 5.0$, $r_0 = 10^{-3}$, $l_0 = 2$, and $\mathcal{T} = 1$ throughout. All quantities are reported in units where $k_B T = 1$.

In the WCA potential, the parameter $\sigma_{ab} = (\sigma_a + \sigma_b)/2$ is the average effective diameter of particles $a$ and $b$. Each particle is randomly assigned a diameter $\sigma_a$ from the set $\{0.01, 0.02, 0.03\}$, a charge $q_a \in \{0.9, 1.0, 1.1\}$, and a diffusion coefficient $D_a$ inversely proportional to $\sigma_a$, with the smallest diameter corresponding to $D = 1$.

Time evolution is simulated using the Euler–Maruyama method with fixed time step $\Delta t = 2 \times 10^{-4}$
according to the update rule
\begin{equation}
\mathbf{x}_a(t + \Delta t) = \mathbf{x}_a(t) + \mu_a\Delta t \, \mathbf{F}_a(t) + \sqrt{2 D_a \Delta t} \, \boldsymbol{\eta}_a,
\end{equation}
where $\boldsymbol{\eta}_a$ is a vector of independent standard normal random variables.

We consider various system sizes $N \in \{1, 2, 5, 10, 20, 50, 60, 80, 100\}$ and run 5 independent simulations for each $N$. Each simulation consists of $3 \times 10^6$ initial transient steps (discarded for analysis) followed by $7.2 \times 10^6$ production steps, during which the full particle configurations are recorded at every time step. The production phase is used to compute time-averaged quantities, which serve as surrogates for ensemble averages. We find that the standard deviation across the five independent simulations is consistently very small, confirming that the transient phase is sufficient for the system to reach a well-defined nonequilibrium (periodic) steady state.

To compute coarse-grained density observables, the simulation box is divided into $5 \times 5$ equal square regions. At each time $t$, the number density $\rho_m(t)$ in {region $m$} is computed by counting particles. For a range of delay steps $n_{\text{obs}}$, {which corresponds to the delay time $\tau=n_{\text{obs}}\Delta t$}, we compute the cross-correlation
\begin{equation}
 C_{nm}^{t,\tau} = \left\langle \rho_n(t) \rho_m(t + \tau) \right\rangle = \left\langle \rho_n(t) \rho_m(t + n_{\text{obs}} \Delta t) \right\rangle,
\end{equation}
where the ensemble average is obtained by taking a long-time (arithmetic) average over the stroboscopic records in the production phase. After discarding the initial transient, the system was assumed to have reached
a periodic steady state, so that stroboscopic samples taken at the same phase of the driving period are statistically equivalent across cycles. We sampled the density records stroboscopically at $t_k=t_0+k\mathcal{T}$, ($k=0,1,\ldots,1440$), choosing $t_0=\mathcal{T}/4$.
Each independent run then provided $1440$ production periods, from which the correlation functions were estimated by arithmetic averaging over the stroboscopic records. The reported values were finally obtained by averaging these run-wise estimates over five independent runs. For the lag times used in the analysis, $\tau=\ell \mathcal{T}$ with positive integer $\ell (\in \mathbb{Z}_{>0})$. With the choice $t_0=\mathcal{T}/4$, the
sinusoidal driving field in Eq.~\eqref{protocol_periodic} is centered at its maximum and is symmetric under time reversal within each observation window: $E(t_0+s)=E(t_0+\tau-s)$ for $0\le s\le \tau$. Since all remaining force components are time independent, the backward protocol is identical to the forward protocol on these windows. Consequently, $C^{\dagger,t_0,\tau}_{mn}=C^{t_0,\tau}_{mn}$. In what follows, we shift the time origin to $t_0$ and write $C^{0,\tau}_{mn}:=C^{t_0,\tau}_{mn}$.

We then use the above estimator to estimate the coarse-grained entropy production as
\begin{equation}
\frac{\mathcal{T}}{\tau} \sigma^{\text{est}}_{[0,\tau]}  = \frac{\mathcal{T}}{2 \, n_{\text{obs}} \Delta t} \sum_{m,n|m \ne n} \left[ C^{0,\tau}_{mn} - C^{0,\tau}_{nm} \right] \ln \left( \frac{C^{0,\tau}_{mn}}{C^{0,\tau}_{nm}} \right).
\end{equation}

We consider a constraint $n_{\rm obs}\Delta t=\tau=\ell \mathcal{T}$. We thus optimize over $\ell$ to obtain the best estimator
\begin{equation}
    \sigma^{\rm est,*}_{[0,\tau]}=\max_{\ell \in \mathbb{Z}_{>0}} \frac{1}{\ell}\sigma^{\text{est}}_{[0,\ell \mathcal{T}]}.
\end{equation}
In practice, evaluating the estimator for all positive integers $\ell$ is
computationally demanding. We therefore restricted the numerical maximization to the finite candidate set $\ell\in \{1,2,\dots,10,15,20,22, 25,30,40\}$ for the results reported in the main text.

The true entropy production per period in the (periodic) steady state is computed by time-averaging the irreversible work done by the external field $\mathbf{F}_a^{\rm ext}$:
\begin{equation}
\frac{\sigma_{[0,\mathcal{T}]} }{\mathcal{T}} = \frac{1}{ \mathcal{T}} \left\langle\int_0^{\mathcal{T}} \sum_{a=1}^N  \mathbf{F}_a^{\text{ext}} \circ \dot{\bf x}_a(t) \, dt\right\rangle,
\end{equation}
where $\circ$ denotes Stratonovich convention, and the notation $\mathbf{F}_a^{\text{ext}} \circ \dot{\bf x}_a(t)$ denotes the scalar product with the Stratonovich convention. Because the external force is $\mathbf{F}_a^{\text{ext}} = q_a E(t) \hat{x}$, the scalar product $\mathbf{F}_a^{\text{ext}} \circ \dot{\bf x}_a(t)$ picks out only the displacement of each particle along the field direction $\hat{x}$. The ensemble average is obtained from the long-time average similarly to above. Since the system is in a periodic steady state, both the net change in system entropy and the net work done by conservative forces vanish over a cycle, so the total entropy production is entirely given by the time-integrated environmental entropy flow due to non-conservative forces.

\section{Relation to previous coarse-graining approaches and origin of the parameter dependence}
\subsection{Deterministic lumping as a special limit of random coarse-graining}

Previous deterministic coarse-graining schemes are recovered by choosing a partition $\{S_m\}_m$ of the fine-grained state space and the indicator kernel $T(m|i)=T_{\rm det}(m|i)$, defined as
\begin{equation}
T_{\rm det}(m|i)=\openone_{S_m}(i).
\end{equation}
Here, $\openone_{S_m}(i)$ is the indicator function of the CG states.
Then each microscopic state contributes to one and only one coarse state, and our general definitions reduce to the standard deterministic lumping formulas. In particular,
\begin{equation}
p_m^{\rm cg}=\sum_{i\in S_m} p_i,
\qquad
\Phi^{\rm det}_{mn}\coloneqq k^{\rm cg}_{mn}p_n^{\rm cg}=
\sum_{\nu}\sum_{i\in S_m}\sum_{j\in S_n}k_{ij}^\nu p_j,
\end{equation}
so the visible entropy production becomes the usual coarse-grained entropy production built from the fluxes between the macrostates.

The two hidden contributions reduce accordingly. The hidden-within-state part becomes
\begin{equation}
\dot\sigma^{\rm inn}=\dot\sigma_{\rm det}^{\rm inn} :=
\sum_m\sum_\nu\sum_{i\in S_m,j\in S_m|i\neq j}
 k_{ij}^\nu p_j\ln\frac{k_{ij}^\nu p_j}{k_{ji}^\nu p_i},
\end{equation}
which is the dissipation from microscopic transitions inside each lump. The transition-mixing part becomes
\begin{equation}
\dot\sigma^{\rm tran}= \dot\sigma_{\rm det}^{\rm tran} :=
\sum_{m,n|m\neq n}\Phi^{\rm det}_{mn}
D_{\rm KL}\!\left[Q^{\rm det}_{mn}\,\middle\|\,Q^{\rm det}_{nm}\right],
\end{equation}
where $D_{\rm KL}\!\left[Q^{\rm det}_{mn}\,\middle\|\,Q^{\rm det}_{nm}\right]:=\sum_{\nu, i,j}Q^{\rm det}_{mn} (\nu, i, j)\ln [Q^{\rm det}_{mn} (\nu, i, j)/Q^{\rm det}_{nm} (\nu, i, j)]$ is the Kullback-Leibler divergence between two probability distributions $Q^{\rm det}_{mn}$ and $Q^{\rm det}_{nm}$, and  $Q^{\rm det}_{mn}$ is defined as
\begin{equation}
Q^{\rm det}_{mn}(\nu,i,j)=
\frac{\openone_{S_m}(i)\openone_{S_n}(j)k_{ij}^\nu p_j}{\Phi^{\rm det}_{mn}}.
\end{equation}
This is exactly the log-sum gap associated with merging several microscopic channels into a single visible transition. Therefore the deterministic stochastic-thermodynamic structure is recovered as a strict special case, in agreement with the coarse-grained stochastic thermodynamics by Esposito \cite{esposito12pre}.

What random coarse-graining adds is the ability to describe situations that deterministic lumping cannot represent. A measurement-error channel $T(m|i)= T(m|i;\varepsilon)$, defined as
\begin{equation}
T(m|i;\varepsilon)=(1-\varepsilon)\delta_{m,i}+\varepsilon\,\delta_{m,P(i)}
\end{equation}
assigns two possible outcomes to the same microscopic state. This is impossible with indicator functions but is the minimal structure needed to model finite-resolution readout, misclassification, or overlapping observables. In this sense, imprecise measurements are not merely a small variant of ordinary lumping. They require genuinely probabilistic coarse outcomes. Recent work on faulty coarse-graining addresses related issues at the trajectory level \cite{Jann25faulty}; our formulation is complementary and works directly at the state/observable level.

The observable interpretation $T(m|i)=O_m^i$ gives a second extension beyond deterministic lumping. Instead of first constructing discrete observed states and then analyzing their dynamics, one can start from any normalized set of nonnegative outputs and obtain a rigorous lower bound written directly in terms of measurable cross-correlations. This turns the asymmetry of observable dynamics into an operational estimator of irreversibility and connects naturally to directed information flow and circulation in observable space \cite{23ohgaprl}. The four-state push-pull example in Sec.~III makes the relation to previous approaches explicit: when $x=0$ or $x=1$, the construction reduces to ordinary deterministic two-state coarse-graining and the bound vanishes; for $0<x<1$, the same measurement channel becomes genuinely random and yields a nonzero lower bound.


\subsection{Why the coarse-grained entropy production depends on $N$ and on the driving strength}

A fixed coarse-graining kernel $T(m|i)$ specifies only which information is retained. It does not determine how much of the total dissipation remains visible after projection. That visibility is controlled by the hidden part
\begin{equation}
\dot{\sigma}-\dot{\sigma}^{\rm cg}
= \dot{\sigma}^{\rm inn}+\dot{\sigma}^{\rm tran}.
\end{equation}
The first hidden term,
\begin{equation}
\dot\sigma^{\rm inn}=
\sum_m\sum_\nu\sum_{i,j|i\neq j}
T(m|i)T(m|j)k_{ij}^\nu p_j\ln\frac{k_{ij}^\nu p_j}{k_{ji}^\nu p_i},
\end{equation}
collects irreversible microscopic transitions that remain indistinguishable at the observed level. In the box-counting application, these are rearrangements of detailed particle configurations or species-resolved motions that are invisible once we keep only the coarse counting output. As the number of particles increases, the number of such unresolved rearrangements typically grows rapidly. A stronger field can also activate more irreversible motion while leaving the same coarse counting signal. There is therefore no reason for $\dot{\sigma}^{\rm inn}$ to stay fixed when $N$ or the driving amplitude is varied at fixed coarse-graining.

The second hidden term,
\begin{equation}
\dot\sigma^{\rm tran}=
\sum_{m,n|m\neq n}\Phi_{mn}
D_{\rm KL}\!\left[Q_{mn}\,\middle\|\,Q_{nm}\right],
\qquad
\Phi_{mn}=k^{\rm cg}_{mn}p_n^{\rm cg},
\end{equation}
measures the information loss coming from mixing many microscopic channels into the same visible transition $n\to m$. It vanishes only in special cases, for example when all microscopic channels contributing to a given visible pair carry the same effective affinity. In general, however, changing $N$ or the driving strength changes the microscopic steady probabilities and the relative weights of different pathways, and therefore changes the Kullback--Leibler gap. This is why the visible fraction of the dissipation can depend strongly on system parameters even though the coarse-graining kernel is kept fixed.

This mechanism is already present in conventional deterministic coarse-graining. In the deterministic limit,
\begin{equation}
\dot\sigma^{\rm cg}= \dot\sigma_{\rm det}^{\rm cg} :=
\sum_{m,n|m\neq n}\Phi^{\rm det}_{mn}
\ln\frac{\Phi^{\rm det}_{mn}}{\Phi^{\rm det}_{nm}},
\qquad
\Phi^{\rm det}_{mn}=
\sum_\nu\sum_{i\in S_m}\sum_{j\in S_n}k_{ij}^\nu p_j,
\end{equation}
and the same hidden transition-mixing term $\dot\sigma^{\rm tran}=\dot\sigma_{\rm det}^{\rm tran}$ remains. Since both the microscopic affinities and their weights depend on the model parameters, the deterministic coarse-grained entropy production is not determined solely by the geometric ``scale'' of the coarse-graining. It also depends on how the microscopic nonequilibrium structure projects onto the retained observables. The full thermodynamic structure is recovered only in special limits, such as rapid equilibration inside each macrostate \cite{esposito12pre}, while more general coarse-graining corrections depend on the hidden currents and on time-scale structure \cite{Busiello19njp,Busiello20prr}.

Therefore the decrease of $\sigma^{\rm est,*}_{[0, \mathcal{T}]}/\sigma_{[0, \mathcal{T}]}$ with $N$ or with the driving strength in our many-body example is not a sign that the coarse-graining is ``inaccurate'' in any pathological sense. It is the generic price of discarded information. Improvement is possible only by using more observables, a finer partition, or particle-resolved information. The point of the present estimator is to work in the experimentally important regime where only counting data are available.

\section{Weak-interaction scaling, enhanced bounds, and the strong-interaction limitation}
\label{sec:weak-enhancement-sm}

In this section, we analyze the $N$-scaling of the estimator in some extreme cases, and provide possible enhancements. The conclusion is summarized as follows.  First, for noninteracting particles whose path probabilities factorize, an enhanced bound can be derived rigorously.  Second, for short-range interacting particles, locality of correlations gives a limiting scaling estimate that explains why deviations from extensive $N$-scaling are common. Third, a collective-trajectory limit shows that no enhancement depending only on $N$ can hold for arbitrary interactions. The numerical behavior in our interacting example lies between this local-correlation limit and the collective strong-coupling limit.

Throughout this section, we consider a steady state, or a periodic steady state with $\tau$ equal to an integer multiple of the driving period.  The one-time box probabilities are then the same at the two endpoints, and the backward two-time correlation is obtained by exchanging the box labels.  We write
\begin{equation}
\sigma_{[0,\tau]}^{\rm est}
:=\sum_{m,n|m<n}\left(C_{mn}-C_{nm}\right)
\ln\frac{C_{mn}}{C_{nm}},
\qquad
C_{mn}:=C_{mn}^{0,\tau}=\left\langle \rho_m(0)\rho_n(\tau)\right\rangle .
\label{eq:Sigma-rho-sm}
\end{equation}
Every term in this expression is nonnegative.  For particle $a$, let $\rho_m^{(a) i}:=\openone_{S_m}(\mathbf{X}_a(i))$ be the indicator that this particle is in box $m$ (for the microscopic configuration $i = \{ \mathbf{x}_b \}_b$) where $\mathbf{X}_a(i)$ is defined as $\mathbf{X}_a(\{ \mathbf{x}_b \}_b) = \mathbf{x}_a$. The measured normalized density and the corresponding one-time probabilities are
\begin{equation}
\begin{aligned}
\rho_m^i&=\frac{1}{N}\sum_{a=1}^N\rho_m^{(a) i},
&
q_m^{(a)}&:=\left\langle\rho_m^{(a)}(0)\right\rangle
=\left\langle\rho_m^{(a)}(\tau)\right\rangle,\\
q_m&:=\left\langle\rho_m(0)\right\rangle
=\left\langle\rho_m(\tau)\right\rangle
=\frac{1}{N}\sum_{a=1}^Nq_m^{(a)},
&
q^*&:=\min_{m:q_m>0}q_m .
\end{aligned}
\label{eq:qstar-sm}
\end{equation}
Thus, all quantities appearing in Eq.~\eqref{eq:Sigma-rho-sm} and in the enhanced bounds below are obtained from the same box-counting data, apart from the known particle number $N$.

\subsection{Rigorous enhancement for independent noninteracting particles}

We first allow the particles to have different single-particle dynamics, but assume that they are statistically independent.  Expanding the product of the measured densities gives the exact identity
\begin{equation}
C_{mn}=\frac{1}{N^2}\sum_{a,b}
\left\langle \rho_m^{(a)}(0)\rho_n^{(b)}(\tau)\right\rangle .
\label{eq:C-particle-expansion-sm}
\end{equation}
For $a\ne b$, independence implies
$\langle\rho_m^{(a)}(0)\rho_n^{(b)}(\tau)\rangle=q_m^{(a)}q_n^{(b)}$.
It is therefore useful to introduce the single-particle two-time correlations and their particle average,
\begin{equation}
\begin{aligned}
C_{mn}^{(a)}&:=\left\langle\rho_m^{(a)}(0)\rho_n^{(a)}(\tau)\right\rangle,
&
\overline C_{mn}^{\rm single}&:=\frac{1}{N}\sum_{a} C_{mn}^{(a)},\\
R_{mn}&:=\frac{1}{N^2}\sum_{a,b|a\ne b}q_m^{(a)}q_n^{(b)},
&
C_{mn}&=R_{mn}+\frac{1}{N}\overline C_{mn}^{\rm single},
\qquad R_{mn}=R_{nm}.
\end{aligned}
\label{eq:independent-decomposition-general-sm}
\end{equation}
The last equality follows by exchanging the dummy particle labels $a$ and $b$.  Equation~\eqref{eq:independent-decomposition-general-sm} makes the origin of the weak signal explicit: correlations formed from two different particles generate a time-symmetric background $R_{mn}$, whereas the time-antisymmetric part is carried only by the same-particle term, whose weight is $1/N$.

Let $\mathbb P_a(\gamma^{(a)})$ and $\mathbb P_a^\dagger(\gamma^{(a) \dagger})$ denote the forward and backward path probabilities of particle $a$. Here, $\gamma^{(a)}\equiv \{\gamma^{(a)}_t \}_{0\leq t \leq \tau} $ and $\gamma^{(a)\dagger}\equiv \{\gamma^{(a)}_{\tau-t} \}_{0\leq t \leq \tau} $ are trajectories over $[0, \tau]$, and $\gamma^{(a)}_t $ is the FG state of the particle $a$ at time t.  For noninteracting particles, both path measures factorize, and relative entropy is additive:
\begin{equation}
\mathbb P(\gamma)=\prod_{a=1}^N\mathbb P_a(\gamma^{(a)}),
\qquad
\mathbb P^\dagger (\gamma^{\dagger})=\prod_{a=1}^N\mathbb P_a^\dagger (\gamma^{(a)\dagger}),
\qquad
\sigma_{[0,\tau]}
=D_{\rm KL}\!\left[\mathbb P\middle\|\mathbb P^\dagger\right]
=\sum_{a=1}^N D_{\rm KL}\!\left[\mathbb P_a\middle\|\mathbb P_a^\dagger\right],
\label{eq:path-factorization-sm}
\end{equation}
where $D_{\rm KL}\!\left[\mathbb P\middle\|\mathbb P^\dagger\right] := \int \mathbb{P} (\gamma) \ln [\mathbb{P} (\gamma)/\mathbb{P}^\dagger (\gamma^{\dagger})] d\gamma$ and $D_{\rm KL}\!\left[\mathbb P_a\middle\|\mathbb P^\dagger_a\right] :=\int \mathbb{P}_a (\gamma^{(a)}) \ln [\mathbb{P}_a (\gamma^{(a)})/\mathbb{P}_a^\dagger (\gamma^{(a)\dagger})] d\gamma^{(a)}  $.
Applying data processing to the two endpoint box labels of each particle, and then applying the log-sum inequality to the sum over $a$, yields
\begin{equation}
\sigma_{[0,\tau]}
\ge \sum_{a}\sum_{m,n|m<n}
\left(C_{mn}^{(a)}-C_{nm}^{(a)}\right)
\ln\frac{C_{mn}^{(a)}}{C_{nm}^{(a)}}
\ge N\sum_{m,n|m<n}
\left(\overline C_{mn}^{\rm single}-\overline C_{nm}^{\rm single}\right)
\ln\frac{\overline C_{mn}^{\rm single}}{\overline C_{nm}^{\rm single}} .
\label{eq:single-particle-logsum-sm}
\end{equation}
For an unordered pair oriented such that $\overline C_{mn}^{\rm single}\ge\overline C_{nm}^{\rm single}$, adding the same nonnegative background to numerator and denominator can only reduce the logarithmic ratio. From $\overline C_{mn}^{\rm single}\ge\overline C_{nm}^{\rm single}$ and $R_{mn}=R_{nm} \ge 0$, we obtain the inequality 
\begin{equation}
0\le
\ln\frac{R_{mn}+\overline C_{mn}^{\rm single}/N}
        {R_{mn}+\overline C_{nm}^{\rm single}/N}
\le
\ln\frac{\overline C_{mn}^{\rm single}}{\overline C_{nm}^{\rm single}},
\label{eq:N2-general-independent-sm}
\end{equation}
Using Eqs.~\eqref{eq:independent-decomposition-general-sm},~\eqref{eq:single-particle-logsum-sm} and~\eqref{eq:N2-general-independent-sm}, we obtain
\begin{equation}
\sigma_{[0,\tau]}\ge N^2\sigma_{[0,\tau]}^{\rm est}.
\end{equation}
This $N^2$ enhancement is therefore a rigorous lower bound for independent particles even when their single-particle dynamics are not identical.  Its derivation uses no approximation and requires no particle-resolved measurement.

A stronger counting-only bound is available when all particles have the same one-time box probabilities, $q_m^{(a)}=q_m$.  This condition holds, in particular, for identical noninteracting particles and for translationally invariant systems with uniform one-time spatial marginals.  In this case, $\overline C_{mn}^{\rm single}$ is the two-time density correlation of the corresponding one-particle system (or the particle-averaged one-particle correlation), which we denote by $C_{mn}^{\rm single}$.  Equation~\eqref{eq:independent-decomposition-general-sm} then reduces to
\begin{equation}
C_{mn}=\frac{N-1}{N} q_mq_n+\frac{1}{N}C_{mn}^{\rm single}.
\label{eq:independent-C-sm}
\end{equation}
The two terms have a direct meaning: with probability $(N-1)/N$ the two density factors select different particles and give the factorized background $q_mq_n$, while with probability $1/N$ they select the same particle and give $C_{mn}^{\rm single}$.  Hence the one-particle correlation is reconstructed exactly from the measured quantities as
\begin{equation}
C_{mn}^{\rm single}=N C_{mn}-(N-1)q_mq_n.
\label{eq:single-particle-reconstruction-sm}
\end{equation}
Substituting this reconstructed correlation into Eq.~\eqref{eq:single-particle-logsum-sm} gives the enhanced lower bound
\begin{equation}
\sigma_{[0,\tau]}
\ge \widehat\sigma_{[0,\tau]}^{\rm ind}
:=N\sum_{m,n| m<n}\left(C_{mn}^{\rm single}-C_{nm}^{\rm single}\right)
\ln\frac{C_{mn}^{\rm single}}{C_{nm}^{\rm single}}.
\label{eq:reconstructed-independent-bound-sm}
\end{equation}

For such uniform cases, we further derive an alternative multiplicative version that can be applied directly to the raw estimator.  Fix a pair with $C_{mn}^{\rm single}>C_{nm}^{\rm single}>0$ and set
\begin{equation}
\lambda:=\frac{1}{N},
\qquad
c_1:=C_{mn}^{\rm single},
\qquad c_2:=C_{nm}^{\rm single},
\qquad r:=q_mq_n,
\qquad
e_{\rm raw}:=(C_{mn}-C_{nm})\ln\frac{C_{mn}}{C_{nm}},
\qquad e_1:=(c_1-c_2)\ln\frac{c_1}{c_2}.
\label{eq:pair-notation-independent-sm}
\end{equation}
By Eq.~\eqref{eq:independent-C-sm},
\begin{equation}
\ln\frac{C_{mn}}{C_{nm}}
=\ln\frac{(1-\lambda)r+\lambda c_1}{(1-\lambda)r+\lambda c_2}
=\int_{c_2}^{c_1}\frac{\lambda\,du}{(1-\lambda)r+\lambda u}.
\label{eq:log-integral-sm}
\end{equation}
Because a two-time joint probability satisfies $u\le\min(q_m,q_n)$ throughout this integral $u \in [c_2, c_1]$, one has $r/u\ge\max(q_m,q_n)\ge q^*$.  Therefore, for $u \in [c_2, c_1]$, 
\begin{equation}
\frac{\lambda u}{(1-\lambda)r+\lambda u}
\le \frac{\lambda}{\lambda+(1-\lambda)q^*},
\end{equation}
and thus,
\begin{equation}
\ln\frac{C_{mn}}{C_{nm}} =\int_{c_2}^{c_1}\frac{\lambda u }{(1-\lambda)r+\lambda u} \frac{du}{u}
\le \frac{\lambda}{\lambda+(1-\lambda)q^*}\ln\frac{c_1}{c_2}.
\label{eq:log-contraction-qstar-sm}
\end{equation}
Using $C_{mn}-C_{nm}=\lambda(c_1-c_2)$ gives
\begin{equation}
e_{\rm raw}
\le \frac{\lambda^2}{\lambda+(1-\lambda)q^*}\,e_1,
\end{equation}
and summing the resulting pairwise inequality gives
\begin{equation}
\sigma_{[0,\tau]}
\ge \widehat\sigma_{[0,\tau]}^{\rm ind}
\ge N^2\bigl[1+(N-1)q^*\bigr]\sigma_{[0,\tau]}^{\rm est},
\label{eq:enhanced-N-sm}
\end{equation}
The case with the opposite orientation follows by exchanging $m$ and $n$, while vanishing probabilities follow by continuity. Let $M$ denote the total number of virtual boxes in the fixed coarse-graining partition. If these $M$ boxes are approximately uniformly occupied and the one-particle correlations  remain $N$-independent, then
\begin{equation}
q_*\simeq\frac{1}{M},
\qquad
N^2\bigl[1+(N-1)q^*\bigr]\sim\frac{N^3}{M},
\qquad
\sigma_{[0,\tau]}^{\rm est}=O(N^{-2}),
\qquad
\widehat\sigma_{[0,\tau]}^{\rm ind}=O(N).
\label{eq:independent-scaling-sm}
\end{equation}
The $O(N^{-2})$ behavior of $\sigma_{[0,\tau]}^{\rm est}$ follows directly from Eq.~\eqref{eq:independent-C-sm}: the antisymmetric part of $C_{mn}$ is $O(N^{-1})$, and the logarithmic asymmetry is also $O(N^{-1})$ in the presence of the $O(1)$ background $q_mq_n$. The $O(N)$ scaling of $\widehat\sigma_{[0,\tau]}^{\rm ind}$ follows directly from Eq.~\eqref{eq:reconstructed-independent-bound-sm}: under these assumptions, the sum in Eq.~\eqref{eq:reconstructed-independent-bound-sm} is $N$-independent and only the overall factor grows with $N$.

\subsection{Possible enhancements for short-range interactions}

For interacting particles, Eq.~\eqref{eq:independent-C-sm} and path additivity no longer hold.  For later use, we define the particle-resolved connected correlations
\begin{equation}
K_{mn}^{ab}(\tau)
:=\left\langle\rho_m^{(a)}(0)\rho_n^{(b)}(\tau)\right\rangle
-q_m^{(a)}q_n^{(b)}.
\label{eq:connected-particle-correlation-sm}
\end{equation}
For any interacting system, this definition gives the exact decomposition
\begin{equation}
C_{mn}=q_mq_n+\frac{1}{N^2}\sum_{a,b} K_{mn}^{ab}(\tau),
\qquad
C_{mn}-C_{nm}
=\frac{1}{N^2}\sum_{a,b}
\left[K_{mn}^{ab}(\tau)-K_{nm}^{ab}(\tau)\right].
\label{eq:connected-decomposition-sm}
\end{equation}
This formula identifies which property is needed for a possible scaling argument. To proceed, we assume that, over the chosen lag $\tau$, the time-antisymmetric connected correlations are summable in particle label: for every relevant box pair there are constants $A_{mn}$ independent of $N$ and a correlation-volume parameter $s_\tau$ such that
\begin{equation}
\sup_{1\le a\le N}\sum_{b=1}^N
\left|K_{mn}^{ab}(\tau)-K_{nm}^{ab}(\tau)\right|
\le A_{mn}s_\tau
\quad\Longrightarrow\quad
\sum_{a,b}
\left|K_{mn}^{ab}(\tau)-K_{nm}^{ab}(\tau)\right|
\le A_{mn}Ns_\tau.
\label{eq:correlation-volume-assumption-sm}
\end{equation}
Equation~\eqref{eq:correlation-volume-assumption-sm} is the assumption used below.  Informally, $s_\tau$ measures how many other particles can carry an appreciable lag-$\tau$ correlation with a given particle; it does not imply a partition into independent clusters.  For finite-range dynamics at fixed density, finite $\tau$, and away from critical or collectively ordered regimes, one may expect $s_\tau=O(1)$, but whether this is valid is model-dependent.

Suppose in addition that the number of boxes is fixed and that, for the box pairs retained in the estimator, $C_{mn}$ and $C_{nm}$ are bounded below by an $N$-independent constant $c_{mn}>0$.  The mean-value theorem gives $|\ln(C_{mn}/C_{nm})|\le |C_{mn}-C_{nm}|/c_{mn}$.  Equations~\eqref{eq:connected-decomposition-sm} and \eqref{eq:correlation-volume-assumption-sm} then imply
\begin{equation}
\left|C_{mn}-C_{nm}\right|\le\frac{A_{mn}s_\tau}{N},
\qquad
0\le\left(C_{mn}-C_{nm}\right)\ln\frac{C_{mn}}{C_{nm}}
\le\frac{A_{mn}^2s_\tau^2}{c_{mn}N^2},
\qquad
\sigma_{[0,\tau]}^{\rm est}=O\!\left(\frac{s_\tau^2}{N^2}\right).
\label{eq:short-range-scaling-sm}
\end{equation}
If, separately, the true entropy production is extensive, $\sigma_{[0,\tau]} \propto N$ with an $N$-independent positive proportionality constant, this gives $\sigma_{[0,\tau]}^{\rm est}/\sigma_{[0,\tau]}=O(s_\tau^2/N^3)$.  This explains why a counting estimator may become increasingly weak even for local interactions.  It is, however, only a scaling estimate. Multiplying the estimator by $N^3/s_\tau^2$ provides a rough order-of-magnitude estimate, but cannot justify a strict lower bound.  Such an enhancement would require additional, model-dependent assumptions.

\subsection{No universal enhancement exists for general interactions}

The preceding enhancement is not a model-free theorem for arbitrary interacting systems.  Without controlling $s_\tau$ or an equivalent tensorization constant, no multiplicative factor depending only on $N$ can be guaranteed. The obstruction is the strong-interaction, or trajectory-collapse, limit.  If attractive interactions approximately lock all particles into one collective trajectory over the observation time, the density field behaves as a single collective coordinate rather than as $N$ independent samples.  In such a case the raw counting estimator may already capture the irreversible motion of this collective coordinate.  Depending on how the collective mobility and driving affinity scale with $N$, the estimator itself may even scale as $O(N)$.  Multiplying it by a universal increasing function of $N$ would then generally overcount the same collective trajectory and can violate the lower-bound property.

Therefore Eq.~\eqref{eq:enhanced-N-sm} is a rigorous enhancement only under the stated independence assumptions.  Short-range interactions can support the scaling estimate in Eq.~\eqref{eq:short-range-scaling-sm}, but for general interactions neither a fixed $N$-scaling nor an enhanced lower bound depending on $N$ alone is available without additional dynamical information.

\section{Additional numerical comparisons}

To benchmark the density-correlation lower bound against a more conventional current-based inference method, we performed an additional comparison in a nonequilibrium steady state.  We used the same heterogeneous interacting-particle model as in the main text, with the same square box of side length $L=5$, periodic boundary conditions, WCA repulsion, finite-range spring interaction, radii $\sigma_a\in\{0.01,0.02,0.03\}$, charges $q_a\in\{0.9,1.0,1.1\}$, diffusion coefficients satisfying the Stokes--Einstein scaling, and the same $5\times 5$ virtual-box partition.  The only change from the periodically driven example in the main text is that the external field is made time independent, $E(t)=E$, so that the system reaches a nonequilibrium steady state.  This steady-state setting is used only for the thermodynamic uncertainty relation (TUR) benchmark, because the standard steady-state TUR does not directly apply to the periodically driven example discussed in the main text. Although there are generalized TURs valid for periodically driven dynamics, those kinds of relations require more detailed information such as the current response to driving period. Therefore, a fair comparison with our estimator is impossible in the periodic driving cases. 

For each saved configuration we record only the normalized box occupations $\rho^i_m(t)=n^i_m(t)/N$,
where $n^i_m(t)$ is the number of particles in virtual box $m$ for configuration $i$. From the same counting time series, we compute the two-time density correlation in the steady state,
\begin{equation}
    C_{mn}^{\tau}:=C_{mn}^{t,\tau}=\left\langle \rho_m(t)\rho_n(t+\tau)\right\rangle .
\end{equation}
In the steady state, $C^{\dagger, t,\tau}_{mn}=C_{mn}^{\tau}$, and our finite-time correlation bound becomes the entropy-production-rate estimator
\begin{equation}
    \dot\sigma^{\rm est}(\tau) := \frac{\sigma^{\rm est}_{[t, t+\tau]}}{\tau}
    =\frac{1}{2\tau}\sum_{m,n| m\neq n}
    \left(C_{mn}^{\tau}-C_{nm}^{\tau}\right)
    \ln\frac{C_{mn}^{\tau}}{C_{nm}^{\tau}} .
    \label{eq:sm-estimator-steady-tur-section}
\end{equation}
We optimize over the sampled lag times and report
\begin{equation}
    \dot{\sigma}^{\rm est, *}:=\max_{\tau}\dot\sigma^{\rm est}(\tau).
\end{equation}

The TUR comparison uses the scalar antisymmetric observable constructed from exactly the same box-counting observables:
\begin{equation}
    J_{\tau}[{\bf X}(t), {\bf X}(t+\tau)]
    :=\sum_{m,n|m<n}\left[
\rho_m^{{\bf X}(t)}\rho_n^{{\bf X}(t+\tau)}-\rho_n^{{\bf X}(t)}\rho_m^{{\bf X}(t+\tau)}\right],
    \label{eq:sm-Jtau-box-current}
\end{equation}
whose mean value in the steady state is $\langle J_{\tau}\rangle=\sum_{m,n|m<n}(\left\langle \rho_m(t)\rho_n(t+\tau)\right\rangle-\left\langle \rho_n(t)\rho_m(t+\tau)\right\rangle)=\sum_{m,n|m<n}(C^{\tau}_{mn}-C_{nm}^{\tau})$. Here, $\rho_m^{\mathbf{X}(t)}$ denotes the particle number density in box $m$ for the microscopic configuration ${\bf X}(t) = \{ {\bf x}_1(t), \dots, {\bf x}_N(t) \}$. 

This observable changes sign under time reversal, i.e. under exchanging $t$ and $t+\tau$.  This type of time-antisymmetric observable satisfies the fluctuation-theorem-based TUR derived in \cite{hasegawa_19tur}. The corresponding TUR-based lower bound is
\begin{equation}
    \dot\sigma^{\rm TUR}(\tau)
    :=\frac{1}{\tau}\ln\!\left[
    1+\frac{2\left\langle J_{\tau}\right\rangle_{\rm}^{2}}
    {{\rm Var}_{}[J_{\tau}]}
    \right]
    \le \dot\sigma,
    \label{eq:sm-tur-box-current}
\end{equation}
where ${\rm Var}[J_{\tau}] = \langle J_{\tau}^2 \rangle - \langle J_{\tau} \rangle^2 $ is the variance in the steady state.
Here, all averages in this TUR-based bound are taken with respect to the steady-state two-time probability density
$P({\bf X}',t+\tau|{\bf X},t)p_{\rm ss}({\bf X})$, where $P({\bf X}',t+\tau|{\bf X},t)$ is the finite-time conditional probability and $p_{\rm ss}({\bf X})$ is the steady-state distribution. It should be noted that the reason we chose this type of the fluctuation-theorem-based TUR and observable $J_{\tau}[{\bf X}(t), {\bf X}(t+\tau)]$ as comparison targets is that $\langle J_{\tau} \rangle$ uses the same information as that used to obtain $\dot{\sigma}^{\rm est}(\tau)$; we did not optimize over all possible TUR-type estimators or observables.

We use the same set of lag times for Eqs.~\eqref{eq:sm-estimator-steady-tur-section} and \eqref{eq:sm-tur-box-current}, and report
\begin{equation}
    \dot\sigma^{\rm TUR,*}:=\max_{\tau}\dot\sigma^{\rm TUR}(\tau).
\end{equation}
The optimization is over the same set of sampled lag times: $\tau\in \{0.05, 0.10, 0.20, 0.50, 1.0, 2.0, 5.0, 10.0, 20.0\}$.
Thus, both methods are constrained to the same experimental information: the virtual-box counting signal.  No tagged-particle trajectories, species-resolved currents, or microscopic forces are used in either estimator. 

\begin{figure}[t]
    \centering
    \begin{minipage}[t]{0.48\linewidth}
        \centering
        \includegraphics[width=\linewidth]{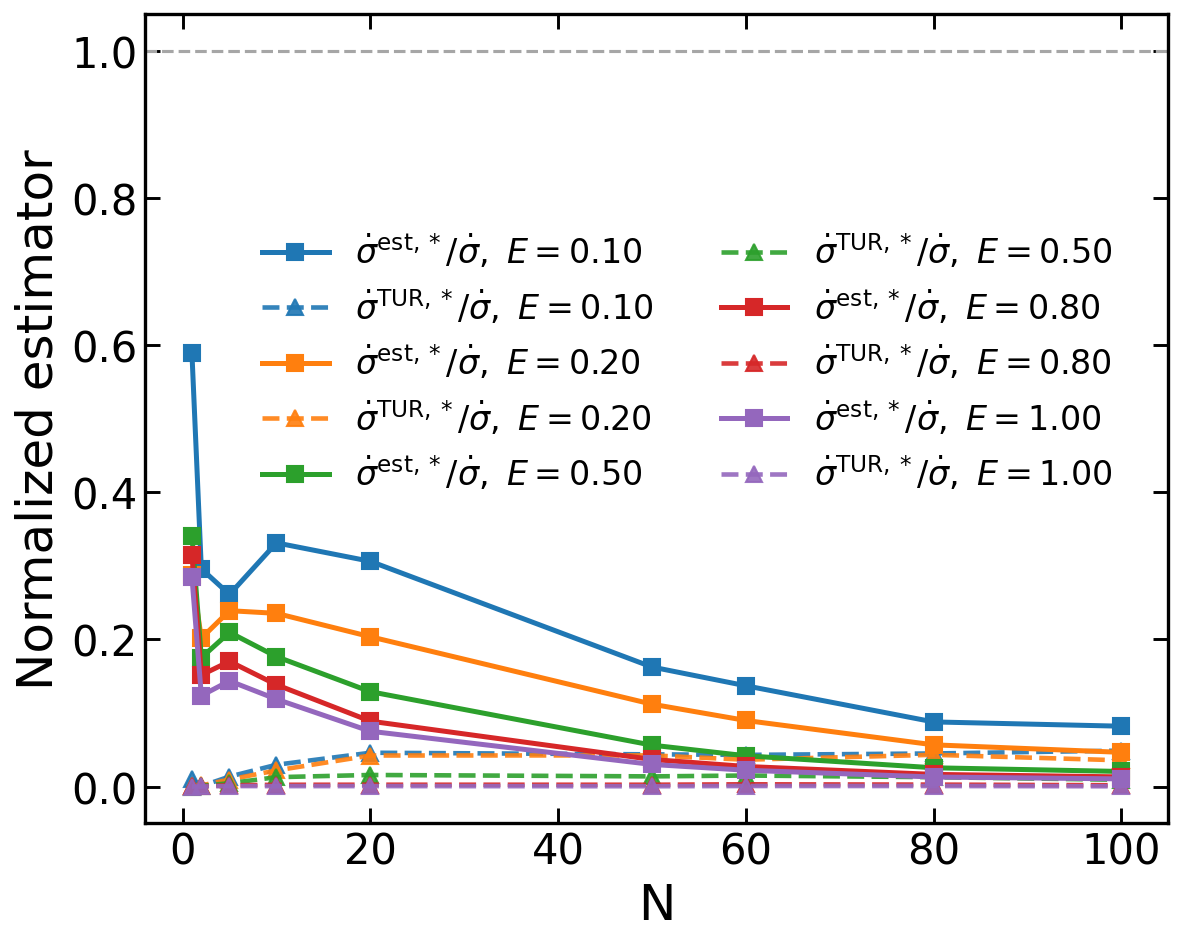}
    \end{minipage}\hfill
    \begin{minipage}[t]{0.49\linewidth}
        \centering
        \includegraphics[width=\linewidth]{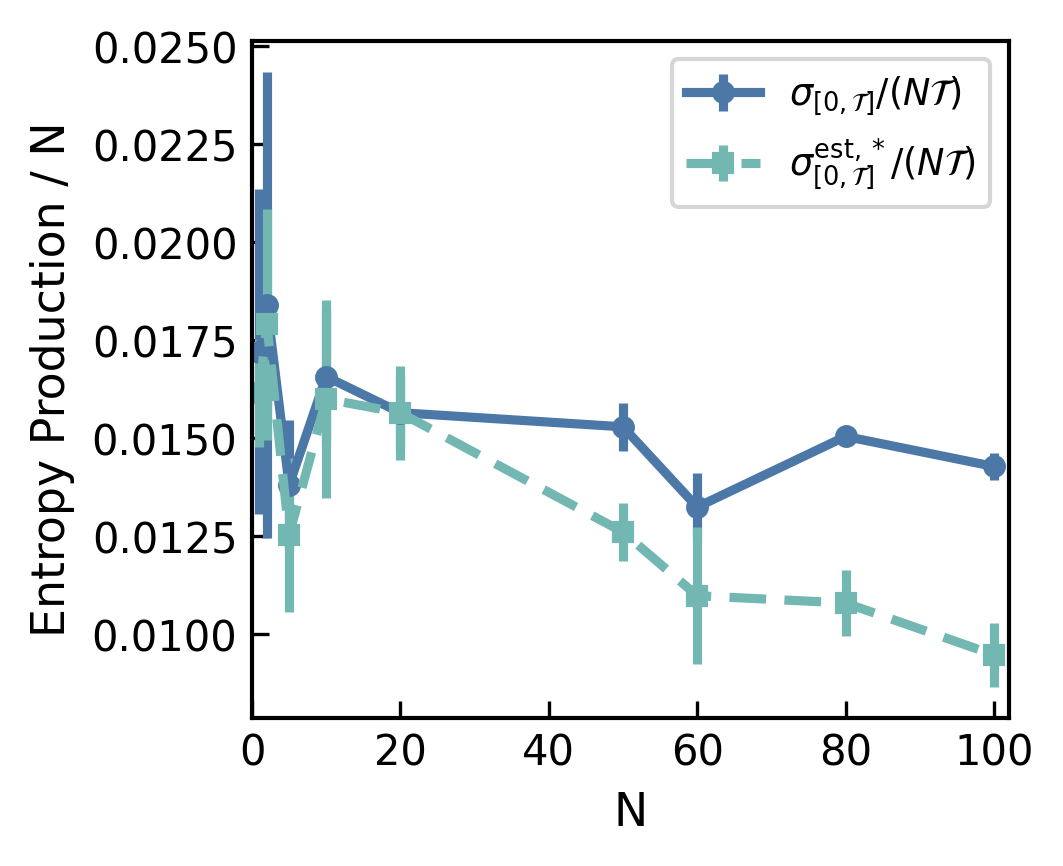}
    \end{minipage}
    \caption{Additional numerical comparisons in the many-body examples. \textbf{Left:} Comparison between the true steady-state entropy production rate $\dot\sigma$, the optimized estimator $\dot\sigma^{\rm est,*}:=\max_{\tau}\,\sigma^{\rm est}_{[0,\tau]}/\tau$, and the optimized TUR-based bound $\dot\sigma^{\rm TUR,*}:=\max_{\tau}\,\dot\sigma^{\rm TUR}(\tau)$ under matched information. The plotted quantities are the ratios $\dot\sigma^{\rm est,*}/\dot\sigma$ (solid lines with square markers) and $\dot\sigma^{\rm TUR,*}/\dot\sigma$ (dashed lines with triangle markers), evaluated from the same virtual-box counting time series in the steady-state variant of the heterogeneous interacting-particle model. Different colors correspond to different driving strengths $E$. \textbf{Right:} Per-particle quantities $\sigma_{[0,\mathcal{T}]}/(N\mathcal{T})$ and $\sigma^{\rm est,*}_{[0,\mathcal{T}]}/(N\mathcal{T})$ for $E_0=E_1=0.1$ in the periodically driven many-body example. The true EPR per particle remains approximately constant, whereas the estimator per particle decreases with $N$, showing explicitly that the extensive $N$-scaling of the fine-grained EPR is not generically preserved after coarse-graining. In both panels, lines are guides to the eye.}
    \label{fig:S3-tur-perparticle}
\end{figure}

The comparison between our estimator and the TUR-based bound is presented in the left panel of Fig. \ref{fig:S3-tur-perparticle}. It is clear that our estimator is stronger in this benchmark under matched information, although the TUR-based bound uses additional information from the variance of the current.

The true steady-state entropy production rate is evaluated from the non-conservative work rate of the external field,
\begin{equation}
    \dot\sigma
    =\left\langle\sum_{a=1}^{N}\mathbf{F}_a^{\text{ext}} \circ \dot {\bf x}_a(t)\right\rangle.
    \label{eq:sm-steady-true-epr}
\end{equation}
Equivalently, in the translationally invariant steady state this work rate can be estimated by the long-time average of the unwrapped particle displacements.

The right panel of Fig.~\ref{fig:S3-tur-perparticle} shows the per-particle quantities $\sigma_{[0,\mathcal{T}]}/(N\mathcal{T})$ and $\sigma^{\rm est,*}_{[0,\mathcal{T}]}/(N\mathcal{T})$ for $E_0=E_1=0.1$ in the periodically driven example. The true EPR per particle stays roughly constant, while the estimator per particle decreases with $N$. This illustrates the point emphasized in the main text: a fixed coarse-graining need not preserve the extensive scaling of the true dissipation, because the fraction of dissipative cycles that remains visible can itself depend on the system parameters.

\bibliographystyle{unsrt}
\bibliography{bibfile}